\newcommand\apjcls{1}
\newcommand\aastexcls{2}
\newcommand\othercls{3}

\newcommand\papercls{\aastexcls}
\documentclass[tighten, times,  preprint2]{aastex6} 

\if\papercls \apjcls
\usepackage{apjfonts}
\else\if\papercls \othercls
\usepackage{epsfig}
\fi\fi
\usepackage{ifthen}
\usepackage{natbib}
\usepackage{amssymb, amsmath}
\usepackage{appendix}
\usepackage{etoolbox}

\if\papercls \apjcls
\newcommand\aas{\ref@jnl{AAS Meeting Abstracts}}
\newcommand\dps{\ref@jnl{AAS/DPS Meeting Abstracts}}
\newcommand\maps{\ref@jnl{MAPS}}
\else\if\papercls \othercls
\usepackage{astjnlabbrev-jh}
\fi\fi

\if\papercls \apjcls
\bibpunct[, ]{(}{)}{,}{a}{}{,}
\else
\bibpunct[, ]{(}{)}{,}{a}{}{,}
\fi

\if\papercls \aastexcls
\hypersetup{citecolor=blue, 
            linkcolor=blue, 
            menucolor=blue, 
            urlcolor=blue}  
\else
\usepackage[
bookmarks=true,           
bookmarksnumbered=true,   
colorlinks=true,          
citecolor=blue,           
linkcolor=blue,           
menucolor=blue,           
urlcolor=blue,            
linkbordercolor={0 0 1},  
pdfborder={0 0 1},
frenchlinks=true]{hyperref}
\fi

\if\papercls \othercls

\else

\fi

\providecommand{\adsurl}[1]{\href{#1}{ADS}}

\makeatletter
\patchcmd{\NAT@citex}
  {\@citea\NAT@hyper@{%
     \NAT@nmfmt{\NAT@nm}%
     \hyper@natlinkbreak{\NAT@aysep\NAT@spacechar}{\@citeb\@extra@b@citeb}%
     \NAT@date}}
  {\@citea\NAT@nmfmt{\NAT@nm}%
   \NAT@aysep\NAT@spacechar\NAT@hyper@{\NAT@date}}{}{}

\patchcmd{\NAT@citex}
  {\@citea\NAT@hyper@{%
     \NAT@nmfmt{\NAT@nm}%
     \hyper@natlinkbreak{\NAT@spacechar\NAT@@open\if*#1*\else#1\NAT@spacechar\fi}%
       {\@citeb\@extra@b@citeb}%
     \NAT@date}}
  {\@citea\NAT@nmfmt{\NAT@nm}%
   \NAT@spacechar\NAT@@open\if*#1*\else#1\NAT@spacechar\fi\NAT@hyper@{\NAT@date}}
  {}{}
\makeatother

\makeatletter
\DeclareRobustCommand{\lowcase}[1]{\@lowcase#1\@nil}
\def\@lowcase#1\@nil{\if\relax#1\relax\else\MakeLowercase{#1}\fi}
\makeatother

\DeclareSymbolFont{UPM}{U}{eur}{m}{n}
\DeclareMathSymbol{\umu}{0}{UPM}{"16}
\let\oldumu=\umu
\renewcommand\umu{\ifmmode\oldumu\else\math{\oldumu}\fi}

\if\papercls \othercls

\else

\fi

\let\oldsim=\sim
\renewcommand\sim{\ifmmode\oldsim\else\math{\oldsim}\fi}
\let\oldpm=\pm
\renewcommand\pm{\ifmmode\oldpm\else\math{\oldpm}\fi}
\newcommand\by{\ifmmode\times\else\math{\times}\fi}

\newcommand\tablebox[1]{\begin{tabular}[t]{@{}l@{}}#1\end{tabular}}
\newbox{\wdbox}
\renewcommand\c{\setbox\wdbox=\hbox{,}\hspace{\wd\wdbox}}
\renewcommand\i{\setbox\wdbox=\hbox{i}\hspace{\wd\wdbox}}

\newcount\timect
\newcount\hourct
\newcount\minct
\newcommand\now{\timect=\time \divide\timect by 60
         \hourct=\timect \multiply\hourct by 60
         \minct=\time \advance\minct by -\hourct
         \number\timect:\ifnum \minct < 10 0\fi\number\minct}

\catcode`@=11

\newcommand\comment[1]{}

\newcommand\commenton{\catcode`\%=14}

\renewcommand\math[1]{$#1$}
\newcommand\mathshifton{\catcode`\$=3}

\let\atab=&
\newcommand\atabon{\catcode`\&=4}

\let\oldmsp=\sp
\let\oldmsb=\sb
\def\sp#1{\ifmmode
           \oldmsp{#1}%
         \else\strut\raise.85ex\hbox{\scriptsize #1}\fi}
\def\sb#1{\ifmmode
           \oldmsb{#1}%
         \else\strut\raise-.54ex\hbox{\scriptsize #1}\fi}
\newbox\@sp
\newbox\@sb
\def\sbp#1#2{\ifmmode%
           \oldmsb{#1}\oldmsp{#2}%
         \else
           \setbox\@sb=\hbox{\sb{#1}}%
           \setbox\@sp=\hbox{\sp{#2}}%
           \rlap{\copy\@sb}\copy\@sp
           \ifdim \wd\@sb >\wd\@sp
             \hskip -\wd\@sp \hskip \wd\@sb
           \fi
        \fi}
\def\msp#1{\ifmmode
           \oldmsp{#1}
         \else \math{\oldmsp{#1}}\fi}
\def\msb#1{\ifmmode
           \oldmsb{#1}
         \else \math{\oldmsb{#1}}\fi}

\def\supon{\catcode`\^=7}

\def\subon{\catcode`\_=8}

\def\supsubon{\supon \subon}

\newcommand\actcharon{\catcode`\~=13}

\newcommand\paramon{\catcode`\#=6}

\comment{And now to turn us totally on and off...}

\newcommand\reservedcharson{ \commenton  \mathshifton  \atabon  \supsubon 
                             \actcharon  \paramon}

\catcode`@=12
\reservedcharson

\if\papercls \apjcls

\else

\fi

\if\papercls \othercls
\else
  \newcommand\inpress{n}
  \if\inpress y
    \received{\today}
    \revised{}
    \accepted{}
    \if\papercls \apjcls
    \slugcomment{}
    \fi
  \else
  \slugcomment{\tablebox{In preparation for {\em ApJ}. DRAFT of {\today}.}}
  \fi
\fi

\newcommand\chisq{\ifmmode{\chi\sp{2}}\else\math{\chi\sp{2}}\fi}
\newcommand\redchisq{\ifmmode{ \chi\sp{2}\sb{\rm red}}
                    \else\math{\chi\sp{2}\sb{\rm red}}\fi}

\usepackage{natbib}
\usepackage{aas_macros}
\def\Civ{C~{\sc iv}}

\begin{document}

\title{An Optically Faint Quasar Survey at $z\sim5$ in the CFHTLS Wide Field:\\ Estimates of the Black Hole Masses and Eddington Ratios}
\author{H. Ikeda\altaffilmark{1}, 
T. Nagao\altaffilmark{2}, 
K. Matsuoka\altaffilmark{3,4,5,11},
 N. Kawakatu\altaffilmark{6}, 
 M. Kajisawa\altaffilmark{2,7},
 M. Akiyama\altaffilmark{8}, 
 T. Miyaji\altaffilmark{9}, 
 and T. Morokuma\altaffilmark{10}
}

\affil{\sp{1}National Astronomical Observatory of Japan, 2-21-1 Osawa, Mitaka, Tokyo 181-8588, Japan\\
        \sp{2}Research Center for Space and Cosmic Evolution, Ehime University, Bunkyo-cho, Matsuyama 790-8577, Japan\\
\sp{3}Department of Astronomy, Kyoto University, Kitashirakawa-Oiwake-cho, Sakyo-ku, Kyoto 606-8502, Japan\\
\sp{4}Dipartimento di Fisica e Astronomia, Universit\'{a} di Firenze, Via G. Sansone 1, I-50019 Sesto Fiorentino, Italy\\
\sp{5}INAF -- Osservatorio Astrofisico di Arcetri, Largo Enrico Fermi 5, I-50125 Firenze, Italy\\
\sp{6}Faculty of Natural Sciences, National Institute of Technology, Kure College, 2-2-11 Agaminami, Kure, Hiroshima 737-8506, Japan\\
\sp{7}Graduate School of Science and Engineering, Ehime University, Bunkyo-cho, Matsuyama 790-8577, Japan\\
\sp{8}Astronomical Institute, Tohoku University, 6-3 Aramaki, Aoba-ku, Sendai 980-8578, Japan\\
\sp{9}Instituto de Astronomia, Universidad Nacional Aut\'onoma de M\'exico, Ensenada, Baja California, Mexico\\
\sp{10}Institute of Astronomy, Graduate School of Science, University of Tokyo, 2-21-1 Osawa, Mitaka 181-0015, Japan}

\email{ikeda.hiroyuki@nao.ac.jp}
\email{$^{11}$JSPS Overseas Research Fellow}
\begin{abstract}
We present the result of our spectroscopic follow-up observation for faint quasar candidates at $z\sim5$ in a part of the Canada-France-Hawaii Telescope Legacy Survey wide field. We select nine photometric candidates and identify three $z\sim5$ faint quasars, one $z\sim4$ faint quasar, and a late-type star. Since two faint quasar spectra show \Civ \ emission line without suffering from a heavy atmospheric absorption, we estimate the black hole mass ($M_{\rm BH}$) and Eddington ratio ($L/L_{\rm Edd}$) of them. The inferred $\rm log \it M_{\rm BH}$ are $9.04\pm0.14$ and $8.53\pm0.20$, respectively. In addition, the inferred $\rm log \it(L/L_{\rm Edd})$ are $-1.00\pm0.15$ and $-0.42\pm0.22$, respectively. If we adopt that $L/L_{\rm Edd}=$ constant or $\propto (1+z)^2$, the seed black hole masses ($M_{\rm seed}$) of our $z\sim5$ faint quasars are expected to be $>10^{5}M_{\odot}$ in most cases. We also compare the observational results with a mass accretion model where angular momentum is lost due to supernova explosions (\citealt{2008ApJ...681...73K}). Accordingly, $M_{\rm BH}$ of the $z\sim5$ faint quasars in our sample can be explained even if $M_{\rm seed}$ is $\sim10^3M_{\odot}$. Since $z\sim6$ luminous qusars and our $z\sim5$ faint quasars are not on the same evolutionary track, $z\sim6$ luminous quasars and our $z\sim5$ quasars are not the same populations but different populations, due to the difference of a period of the mass supply from host galaxies. Furthermore, we confirm that one can explain $M_{\rm BH}$ of $z\sim6$ luminous quasars and our $z\sim5$ faint quasars even if their seed black holes of them are formed at $z\sim7$.
 \end{abstract}
 

\section{INTRODUCTION}

It is one of the important issues to elucidate the formation and the evolution of quasars, because this issue is tightly coupled with the physics of the formation and evolution of supermassive black holes (SMBHs). To clarify this issue, large quasar surveys have been performed up to $z\sim7$ by the 2dF QSO Redshift Survey (e.g., \citealt{2001MNRAS.322L..29C}), the Sloan Digital Sky Survey (SDSS) Quasar Survey \citep[e.g.,][]{2006AJ....131.1203F,2006AJ....131.2766R, 2008AJ....135.1057J, 2009AJ....138..305J,2015AJ....149..188J,2016ApJ...833..222J}, the Canada-France High-$z$ Quasar Survey (CFHQS; \citealt{2007AJ....134.2435W}; \citealt{2009AJ....137.3541W}; \citealt{2010AJ....139..906W}), the Panoramic Survey Telescope And Rapid Response System (Pan-STARRS; \citealt{2014AJ....148...14B,2015ApJ...801L..11V,2016ApJS..227...11B,2017MNRAS.466.4568T}), the Dark Energy Survey (DES; \citealt{2015MNRAS.454.3952R,2017MNRAS.468.4702R,2017ApJ...839...27W}), the Subaru High-z Exploration of Low-luminosity Quasar Survey (SHELLQs; \citealt{2016ApJ...828...26M,2017arXiv170405854M}), and some other quasar surveys (e.g., \citealt{2008AJ....136..954G,2010ApJ...723..184K,2010RAA....10..745W,2011Natur.474..616M,2011ApJ...728L..25I,2012ApJ...756..160I,2012ApJ...755..169M,2013A&A...557A..78M,2013ApJ...779...24V,2015MNRAS.451L..16C,2016AJ....151...24A,2016ApJ...819...24W,2016JKAS...49...25J,2017AJ....153..184Y,2017arXiv170608454J}).
Then the quasar luminosity functions (QLF) are derived up to $z\sim6$ (e.g., \citealt{2009MNRAS.399.1755C,2015ApJ...798...28K,2016ApJ...829...33Y,2017arXiv170405996A}). 

Several studies have reported that the slope of the QLFs is different between the lower and higher-luminosity ranges, and the QLFs are generally fitted by the double power-law function (e.g., \citealt{1988MNRAS.235..935B}).
\cite{2009MNRAS.399.1755C} investigated the redshift evolution of the quasar space density and they confirmed that the quasar space density of faint quasars peaks at a lower redshift than that of more luminous quasars (see also \citealt{2011ApJ...728L..25I,2012ApJ...756..160I,2016ApJ...832..208N}). 
This is known as the active galactic nuclei (AGN) downsizing evolution. The AGN downsizing has been also reported by X-ray AGN surveys (\citealt{2003ApJ...598..886U,2014ApJ...786..104U,2005A&A...441..417H,2015ApJ...804..104M,2015MNRAS.451.1892A,2016A&A...587A.142F,2016A&A...590A..80R}). However, the physical origin of the AGN downsizing has not been clarified, that makes high-$z$ faint
quasar surveys are more important \citep[see][for theoretical works on the AGN downsizing evolution]{2012MNRAS.419.2797F,2014ApJ...794...69E}.

Measuring the masses and Eddington ratios of SMBHs is also useful to investigate the formation and the evolution of quasars.
The masses and Eddington ratios of SMBHs can be measured from the single-epoch virial estimators (e.g., \citealt{2011ApJ...730....7T,2012ApJ...753..125S,2014ApJ...790..145D,2011ApJS..194...45S,2012ApJ...761..143N,2013ApJ...771...64M,2014ApJ...795L..29Y,2015ApJ...806..109J,2015ApJ...815..128K,2016ApJ...825....4T,2016PASJ...68....1S}).
A number of studies for the growth history of the SMBHs have been reported so far (e.g., \citealt{2007ApJ...671.1256N,2010ApJ...719.1315K,2011ApJ...730....7T,2016PASJ...68....1S}).
\cite{2007ApJ...671.1256N} investigated the black hole growth for quasars at $z\sim2.3-3.4$. They found that the required growth time for many quasars is longer than the age of the universe for the quasar redshift, suggesting that their $z\sim2.3-3.4$ quasars must have had at least one previous episode of faster growth at higher redshift.
\cite{2011ApJ...730....7T} estimated the evolutionary tracks of the SMBHs in 40 luminous SDSS quasars at $z\sim4.8$, and they mentioned that $\sim40\%$ of $z\sim4.8$ luminous quasars could have been growing up from the stellar black hole mass.
However, most of these studies focus on luminous quasars, due to the lack of the faint quasars at $z>4$.
Consequently it is not understood how faint quasars evolved at high redshift. 
As the number density of faint quasars is much higher than that of luminous quasars, the whole picture of SMBH evolution cannot be understood without understanding the growth history of faint quasars at such high redshifts.

Some $z\sim5$ faint quasar surveys have been carried out so far (e.g., \citealt{2012ApJ...756..160I,2013ApJ...768..105M,2013A&A...557A..78M}).
\cite{2013ApJ...768..105M} discovered 71 $z\sim5$ quasars at $i'<22.0$ and derived the faint side of the QLF at $z\sim5$. While they construct a large sample of faint quasars at $z\sim5$, the achieved signal-to-noise ratio of the spectra is not sufficient to derive the black hole mass.
As for fainter quasar survey at $z\sim5$,
\cite{2012ApJ...756..160I} reported their faint quasar survey for $z\sim5$ quasars with $22<i'<24$ in the COSMOS field 
(1.64 deg$^2$) and gave only the upper limits on the quasar number density
(\citealt{2012ApJ...756..160I}). This is not due to the limiting flux of spectroscopic
observations but simply due to too narrow area of the survey field. Actually
the inferred upper limit on the quasar number density is close to the extrapolated
number density from lower redshifts, suggesting that quasar surveys for
somewhat wider area will find some faint quasars at $z\sim5$. Therefore we focus on the public
database of CFHT legacy survey (CFHTLS; \citealt{2012AJ....143...38G}). Among the CFHTLS-Wide fields ($\sim145$ 
deg$^2$), we specifically focus on a $\sim6$ deg$^2$ area that is covered also by the United Kingdom Infrared Telescope (UKIRT) Infrared Deep Sky Survey (UKIDSS; \citealt{2007MNRAS.379.1599L})-Deep Extragalactic Survey (DXS) to select faint quasars effectively. 

This paper is organized as follows.
In Section 2, we describe our photometric survey for faint quasars at $z\sim5$.
In Section 3, we report the results of follow-up spectroscopic observations, and also the derived black hole mass and Eddington ratio.
In Sections 4 and 5, we give our discussion and summary. In this paper, we adopt a $\Lambda$CDM cosmology with $\Omega_m$ = 0.3, $\Omega_{\Lambda}$ = 0.7, and a Hubble constant of $H_0$ = 70 km s$^{-1}$ Mpc$^{-1}$ (e.g., \citealt{2003ApJS..148..175S}).
All magnitudes and colors are given in the AB system (\citealt{1983ApJ...266..713O}). All magnitudes have been corrected for the Galactic extinction (\citealt{2011ApJ...737..103S}).

\section{The Sample }
\subsection{The Canada-France-Hawaii Telescope Legacy Survey}
The CFHTLS consists of the deep, wide, and very wide surveys which have been performed by MegaPrime/MegaCam (\citealt{2003SPIE.4841...72B}).  We use the photometric data of the wide survey among them. The photometric data were obtained through the $u^{*}$-, $g'$-, $r'$-, $i'$-, and $z'$-band filters. The whole survey field is $\sim 145$ deg$^{2}$. The limiting magnitudes for point sources at 50 $\%$ completeness are $u^{*}=26.0$, $g'=26.5$, $r'=25.9$, $i'=25.7$, and $z'=24.6$ respectively (\citealt{2012AJ....143...38G}). The CFHTLS unified wide catalogs (\citealt{2012AJ....143...38G}) have been produced by MegaPipe (\citealt{2008PASP..120..212G}). There are the $u^{*}$-, $g'$-, $r'$-, $i'$-, and $z'$-band selected catalogs. Since $z\sim5$ quasars can be selected by the $r'$-dropout method, the $u^{*}$-, $g'$-, and $r'$-band selected catalogs are inadequate to select $z\sim5$ quasars. Moreover, the limiting magnitude and seeing of the $i'$-band are fainter and better than those of the $z'$-band. Therefore we use the $i'$-band selected catalogs to select $z\sim5$ faint quasar candidates. Since the number density of quasars at high redshift is quite low, it is very useful to search for quasars by utilizing such a wide photometric catalog.  
\subsection{The United Kingdom Infrared Telescope Infrared Deep Sky Survey}
The UKIDSS consists of the Large Area Survey, the Galactic Clusters Survey, the Galactic Plane Survey, the Deep Extragalactic Survey, and the Ultra Deep Survey. These surveys have been conducted with the Wide Field Camera (WFCAM; \citealt{2007A&A...467..777C}) on the 3.8-m UKIRT. UKIDSS uses a photometric system described in \cite{2006MNRAS.367..454H}. The pipeline processing and science archive are described in Irwin et al (in prep) and \cite{2008MNRAS.384..637H}. We utilizise the UKIDSS DR 10 in this work.
 The area and 5-sigma depths of the Deep Extragalactic Survey, that is focused on in this work, are 35 deg$^2$, $J\sim23.4$ ($J_{\rm vega}\sim22.5$), and $K\sim22.9$ ($K_{\rm vega}\sim21.0$), respectively.
\begin{figure*}[!t]
\begin{center}
\includegraphics[bb= 0 0 700 498,clip,width=18.0cm]{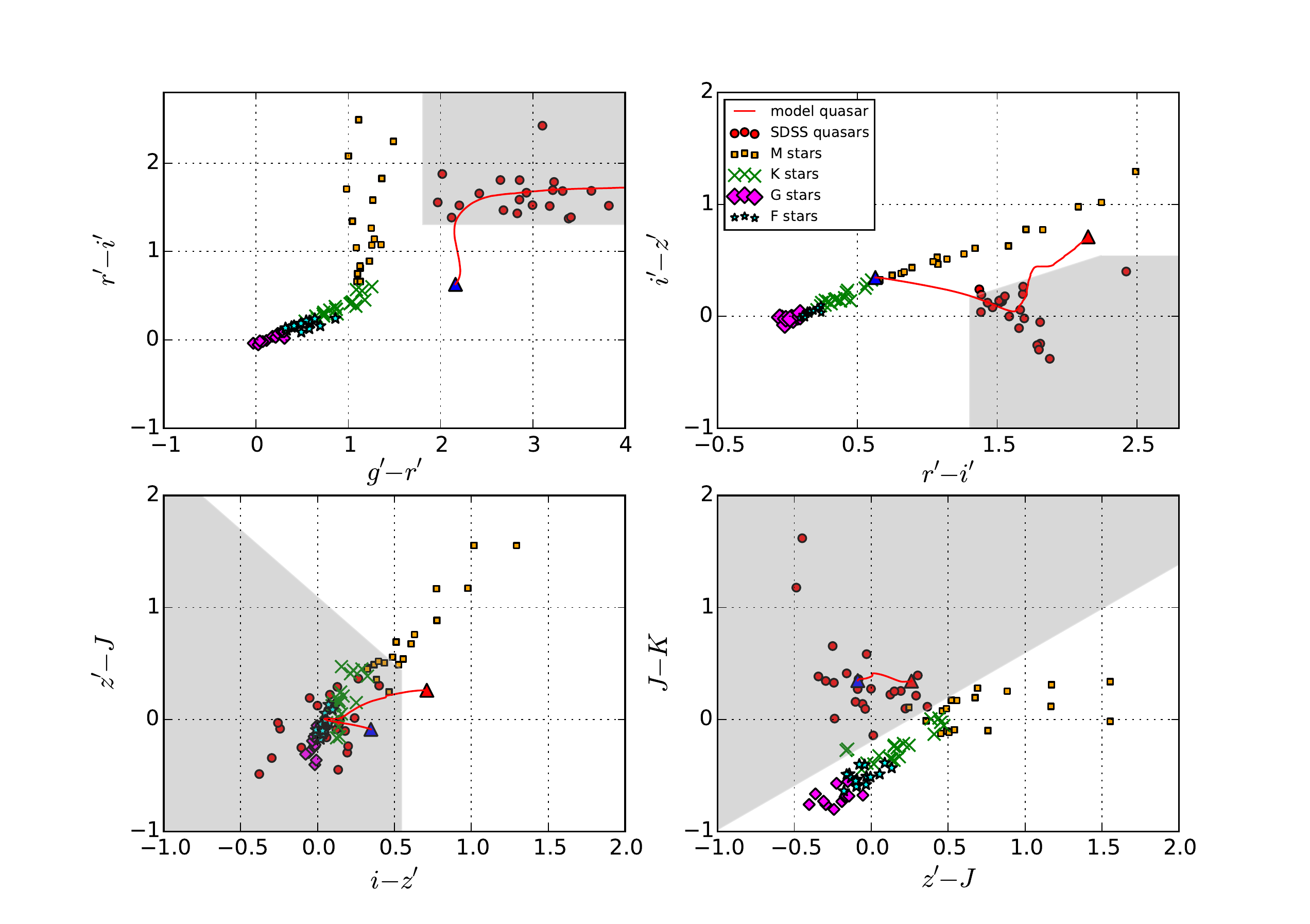}
\caption{Two-color diagrams of $r'-i'$ vs. $g'-r'$, $i'-z'$ vs. $r'-i'$, $z'-J$ vs. $i'-z'$, and $J-K$ vs. $z'-J$. Orange squares, green crosses, cyan stars, and purple diamonds show the colors of M-, K-, G-, and F-type stars which are calculated by utilizing the star spectra library (\citealt{1998PASP..110..863P}), respectively. Red circles and lines show the colors of SDSS quasars and color track of model quasars at $4.5<z<5.5$, respectively. Blue and red triangles show the colors of model quasars at $z=4.5$ and 5.5, respectively. Gray shaded regions in the two-color diagrams show the $z\sim5$ faint quasar candidates regions in this paper.}  
\end{center}
\end{figure*}
\subsection{Selection Criteria for Faint Quasar Candidates at $z\sim5$}
In order to determine the selection criteria of quasars at $z\sim5$ with high completeness and low contamination rate, we check colors of spectroscopically confirmed SDSS quasars at $4.8<z<5.3$ on the CFHTLS photometric systems by utilizing the SDSS quasar catalog data release 12 (\citealt{2017A&A...597A..79P}). 
Since the response curve of the MegaCam filters is slightly different from that of the SDSS filters (see Figure 1 of \citealt{2008PASP..120..212G}),
we calculate the $g'$-, $r'$-, $i'$-, and $z'$-band magnitude of the SDSS quasars by the following relations (\citealt{2008PASP..120..212G}):
\begin{eqnarray}
g' = g_{\rm SDSS} - 0.153 (g_{\rm SDSS} - r_{\rm SDSS}),\\
r' = r_{\rm SDSS} - 0.024 (g_{\rm SDSS} - r_{\rm SDSS}),\\
i' = i_{\rm SDSS} - 0.085 (r_{\rm SDSS} - i_{\rm SDSS}),
\end{eqnarray}
and,
\begin{equation}
z' = z_{\rm SDSS} + 0.074 (i_{\rm SDSS} - z_{\rm SDSS}).
\end{equation}
Using the equations (1) -- (4), we calculate the $g'-r'$, $r'-i'$, and $i'-z'$ of the SDSS quasars.
 Since the SDSS quasar catalog data release 12 includes the information of the $J$-band and $K$-band magnitudes from UKIDSS, we calculate the $z'-J$ and $J-K$ of the SDSS quasars by utilizing them. The colors of stars are also calculated by utilizing the star spectra library (\citealt{1998PASP..110..863P}) to prevent the contamination by stars in selecting $z\sim5$ quasar candidates. In addition, the model quasar is generated by the same method of \cite{2012ApJ...756..160I} and the colors of this model quasar are calculated. We then check some two-color diagrams (Figure 1). As seen in Figure 1, it is useful to select quasars at $z\sim5$ by utilizing these two-color diagrams. Therefore we determine the selection criteria for the faint quasar candidates at $z\sim5$ based on the calculated colors of SDSS quasars and stars on these two-color diagrams. The detailed description of the selection criteria for the faint quasar candidates at $z\sim5$ is provided later in this section.

\begin{figure*}[!t]
\begin{center}
\includegraphics[bb= 0 0 700 488,clip,width=18cm]{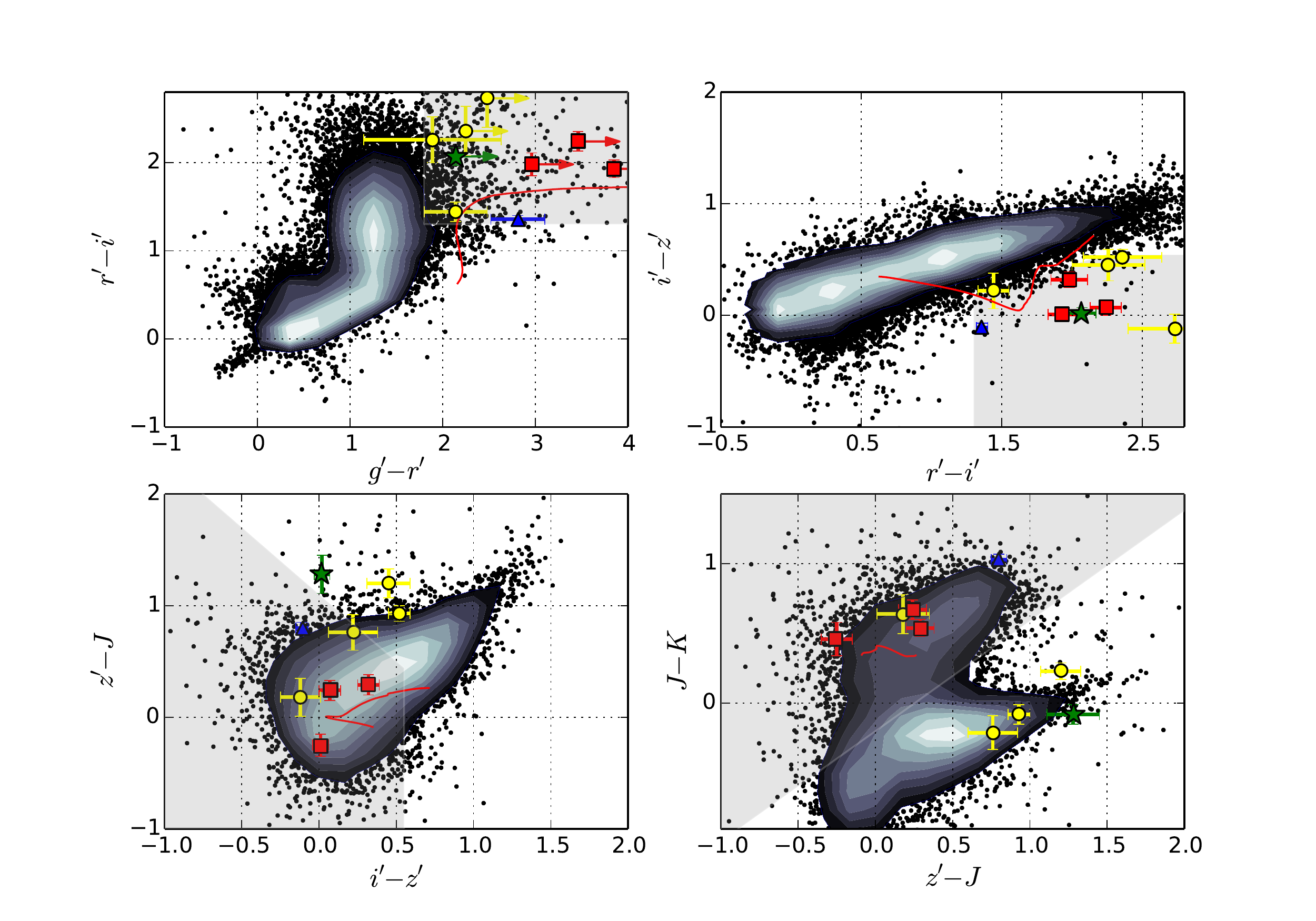}
\caption{Two-color diagrams of $r'-i'$ vs. $g'-r'$, $i'-z'$ vs. $r'-i'$, $z'-J$ vs. $i'-z'$, and $J-K$ vs. $z'-J$. Red lines are the same as in Figure 1. Black points show the colors of point sources which satisfy equations (5) and (7). Red squares, blue triangles, and green stars show the colors of spectroscopically confirmed $z\sim5$ faint quasars, $z\sim4$ faint quasars, and a star in our spectroscopic sample, respectively. Yellow filled circles show the colors of $z\sim5$ faint quasar candidates which we have not yet performed the spectroscopic follow-up observations. Gray shaded regions in the two-color diagrams show the $z\sim5$ faint quasar candidates regions. To make contours, the astroML python package is used (\citealt{astroML,astroMLText}). }  
\end{center}
\end{figure*}
\begin{table*}[hbt]
\begin{center}

\caption{Photometry and spectroscopic follow-up results of faint quasar candidates at $z\sim5$}
\begin{tabular}{c@{\hspace{0.1cm}}c@{\hspace{0.1cm}}c@{\hspace{0.15cm}}c@{\hspace{0.15cm}}c@{\hspace{0.15cm}}c@{\hspace{0.15cm}}c@{\hspace{0.1cm}}c@{\hspace{0.05cm}}c@{\hspace{0.05cm}}c@{\hspace{0.1cm}}c@{\hspace{0.1cm}}c@{\hspace{0.1cm}}c@{\hspace{0.1cm}}c@{\hspace{0.1cm}}c@{\hspace{0.1cm}}c@{\hspace{0.1cm}}c@{\hspace{0.01cm}}c} \hline\hline
       ID&  $i' (\rm MAG\_AUTO)$ &$g'-r'$ &$r'-i'$ & $i'-z'$& $z'-J'$& $J'-K'$& Exp. Time &&  $z_{\rm sp}^{a}$&&Type& \\
        && &&&&&(min)& & && & \\ \hline
J221141.01+001118.92  & $22.01\pm0.02$ & $>3.46$ & $2.24\pm0.11$ & $0.07\pm0.07$ & $0.24\pm0.09$ & $0.67\pm0.07$ &60&&5.23  &&non-BAL QSO$^{b}$&\\ 
J221520.22-000908.39 & $22.25\pm0.02$ & $>2.96$ & $1.98\pm0.13$ & $0.32\pm0.07$ & $0.29\pm0.09$ & $0.54\pm0.06$ & 60 &&5.28 &&BAL QSO$^{c}$&\\ 
J221941.90+001256.20 & $21.56\pm0.01$ & $2.81\pm0.29$ & $1.36\pm0.04$ & $-0.11\pm0.04$ & $0.80\pm0.05$ & $1.03\pm0.04$ & 60&&4.29 && FeLoBAL QSO$^{d}$&\\ 
J222216.02-000405.66 & $21.95\pm0.01$ & $>3.85$ & $1.93\pm0.10$ & $0.01\pm0.05$ & $-0.25\pm0.10$ & $0.46\pm0.12$ & 45 &&4.94&&non-BAL QSO$^{b}$&\\ 
J221653.11+000932.62 & $22.60\pm0.02$ & $>2.14$ & $2.07\pm0.10$ & $0.02\pm0.05$ & $1.28\pm0.17$ & $-0.08\pm0.07$ & 45 &&--&&Star&\\ 

J221254.03+003613.14 & $22.80\pm0.03$ & $2.14\pm0.34$ & $1.44\pm0.11$ & $0.22\pm0.16$ & $0.76\pm0.16$ & $-0.21\pm0.12$ & -- &&--&&--&\\ 
J221309.67-002428.09 & $22.61\pm0.03$ & $>2.48$ & $2.73\pm0.33$ & $-0.12\pm0.13$ & $0.18\pm0.17$ & $0.64\pm0.14$ & -- &&--&&--&\\ 
J221451.49-000220.52 & $22.98\pm0.04$ & $1.89\pm0.74$ & $2.26\pm0.26$ & $0.45\pm0.14$ & $1.20\pm0.13$ & $0.23\pm0.06$ & -- &&--&&--&\\ 
J222205.13+001721.51 & $22.71\pm0.03$ & $>2.25$ & $2.36\pm0.28$ & $0.52\pm0.07$ & $0.93\pm0.07$ & $-0.08\pm0.07$ & -- &&--&&--&\\ 
\hline
              \end{tabular}              
      \end{center}
         $^{a}$ Spectroscopic redshift.\\
         $^{b}$ \cite{2013ApJ...768..105M} also identified them without C {\sc iv} emission lines, due to the wavelength coverage.\\
         $^{c}$ The new faint quasar discovered by our survey.\\ 
         $^{d}$ \cite{2013ApJ...768..105M} also identified it.
\end{table*}
To select faint quasars at $z \sim5$ in the CFHTLS wide field, we adopt the following selection criteria:
\begin{eqnarray}
     21.0 < i' (\rm MAG\_AUTO)<23.0,\\
     i'-z'< 0.40(r'-i')-0.35, \\
     R_{\rm hl}<R_{\rm peak} + 3\sigma,\\
      u^*\geq 2\sigma , \\
        g'-r' \geq 1.8, \\
        1.3<r'-i'<2.8,
            \end{eqnarray}  
and,
\begin{equation}
   i'-z'<0.55,\\
    \end{equation}
 where $R_{\rm hl}$ and $R_{\rm peak}$ are the half-light radius of objects and the peak of $R_{\rm hl}$ distributions for all objects at $21.0 < i' (\rm MAG\_AUTO)<23.0$ in the CFHTLS wide field, respectively. Although $R_{\rm hl}$ which satisfies the criterion (7) is not the same in each field, the typical $R_{\rm hl}$ and $3\sigma$ are $\sim0\farcs47$ and $\sim0\farcs1$, respectively. The criteria (5) and (6) are utilized to select faint quasars without large numbers of contaminants such as stars. Here extended objects are excluded to remove
contaminations such as Lyman-break galaxies (LBGs) at similar redshift and elliptical galaxies at a lower redshift, by utilizing the criterion (7) as the
manner described in \cite{2009A&A...500..981C}. In order to eliminate the low-redshift objects, we add the selection criteria (8), (9), (10), and (11). Utilizing above selection criteria, we select nine faint quasar candidates at $z\sim5$. The photometric properties of all candidates are summarized in Table 1.
 
 It is reported by \cite{2013ApJ...768..105M} that the additional color criteria utilizing near-infrared photometric data are also useful to remove contaminating Galactic stars from the optical color-selected candidates of quasars at $z\sim5$. We investigate the near-infrared colors of SDSS quasars to define the color criteria to remove the stars.
As a result, we confirm that quasars and stars are distinguished effectively by the following equations (see also lower panels of Figure 1):
 \begin{equation}
 z'-J< -1.20(i'-z')+1.10, 
   \end{equation}  
and,
  \begin{equation}  
  J-K> 0.79(z'-J)-0.20. 
  \end{equation}  
  In order to calculate the near-infrared colors ($z'-J$ and $J-K$) of quasar candidates at $z\sim5$ in our survey, we use the UKIDSS-DXS catalog. We confirm that all of the $z\sim5$ faint quasar candidates selected by the equations (5)--(11) are detected in the $J$ and $K$ band.
Then we select five faint quasar candidates at $z\sim5$ by utilizing equations (5)--(13).
Figure 2 shows some two-color diagrams ($r'-i'$ vs. $g'-r'$, $i'-z'$ vs. $r'-i'$, $z'-J$ vs. $i'-z'$, and $J-K$ vs. $z'-J$), where objects which satisfy the criterion (7) down
to $i^\prime = 23.0$ are plotted. We define our survey
limit to $i^\prime = 23.0$ because the number density of LBGs is much higher
than that of quasars at this magnitude (\citealt{2003PASJ...55..415I,2008ApJ...679..269Y}). 
To check how the additional near-infrared criteria are useful to select $z\sim5$ faint quasars, 
we choose the spectroscopic follow-up candidates in faint quasar candidates selected by equations (5)--(11).

\begin{figure*}[t!]
\begin{center}
\includegraphics[bb= 0 0 690 449,clip,width=20.0cm]{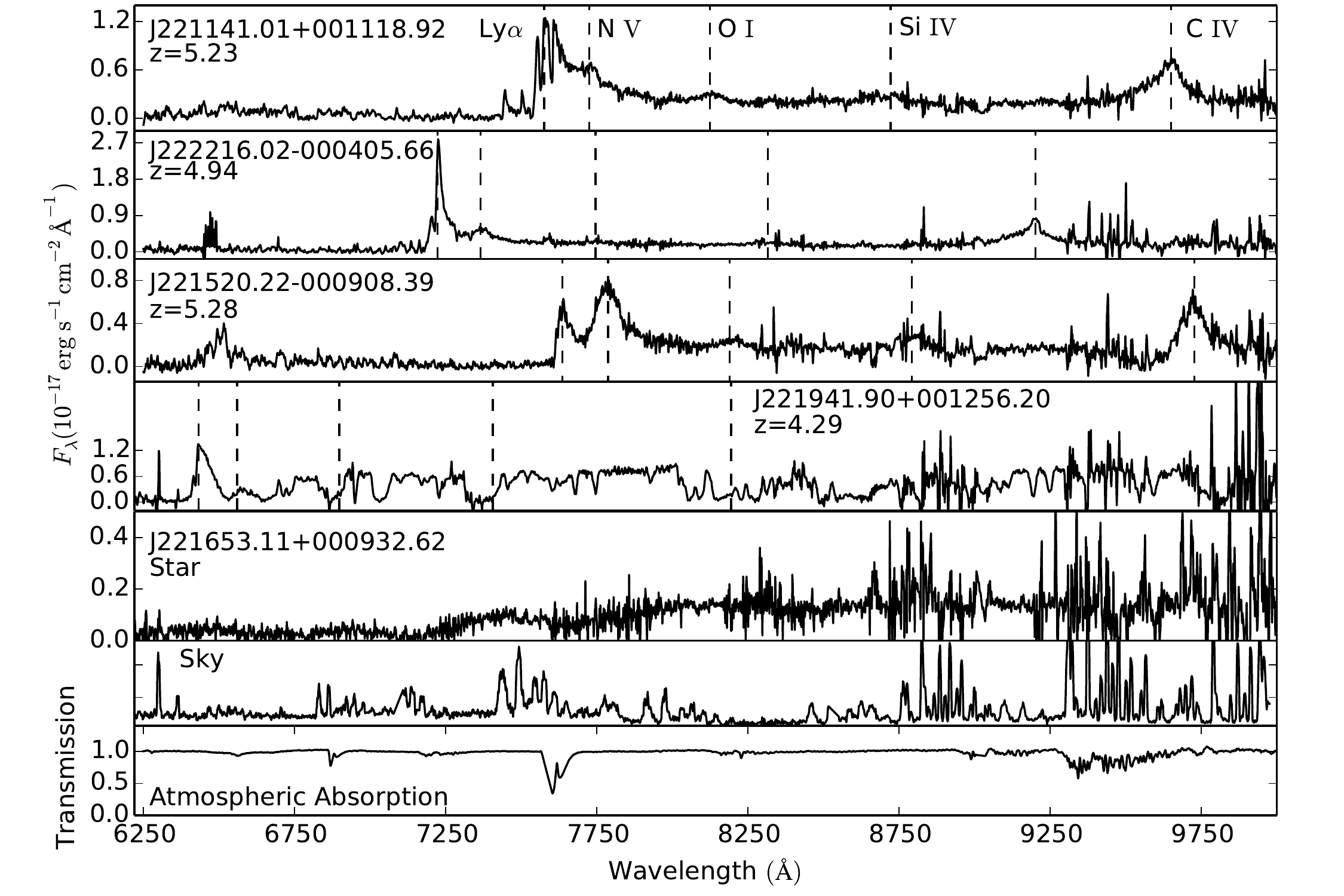}
\caption{Reduced spectra of $z\sim4-5$ faint quasars, the late-type star, the typical sky spectrum, and the atmospheric absorption which is generated by utilizing the spectrum of a standard star (EG131). The dotted lines show the expected wavelengths of quasar emission lines: Ly$\alpha$ $\lambda1216$, {N {\sc v} $\lambda1240$}, {O {\sc i} $\lambda1304$}, {Si {\sc iv} $\lambda1400$}, and {C {\sc iv} $\lambda1549$}.}  
\end{center}
\end{figure*}

\begin{figure*}[!t]
\begin{center}
\includegraphics[bb= 0 0 520 389,clip,width=15.0cm]{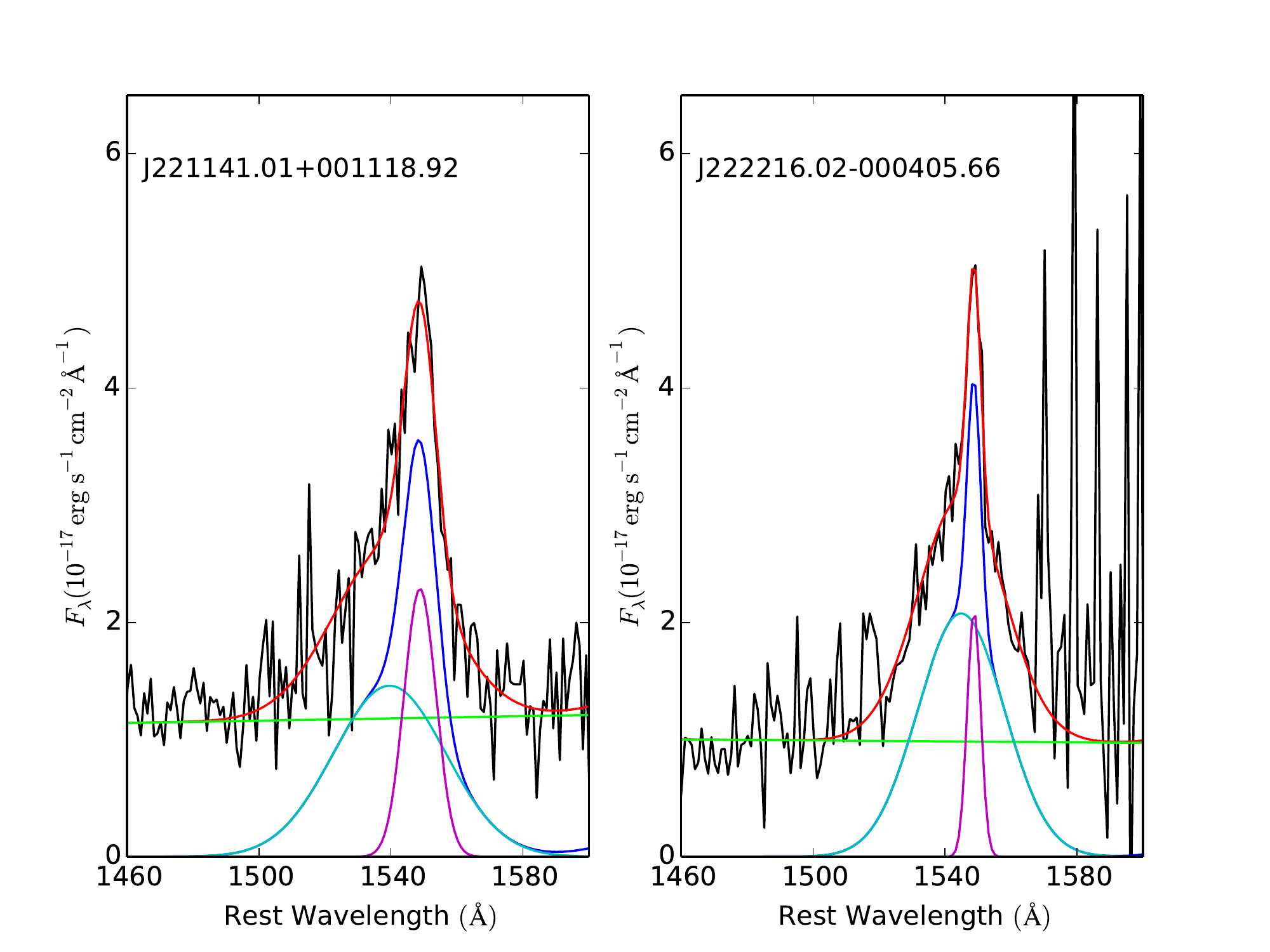}
\caption{Spectral line fitting for the {C {\sc iv} $\lambda1549$} line. Red, blue, light blue, magenta, and green lines show the fitting results of the best fit model, the broad and narrow components of the {C {\sc iv}} emission line, the components of the broad line, the components of the narrow line, and the continuum, respectively.}  
\end{center}
\end{figure*}

\begin{figure*}[!t]
\begin{center}
\includegraphics[bb= 0 0 920 490,clip,width=24cm]{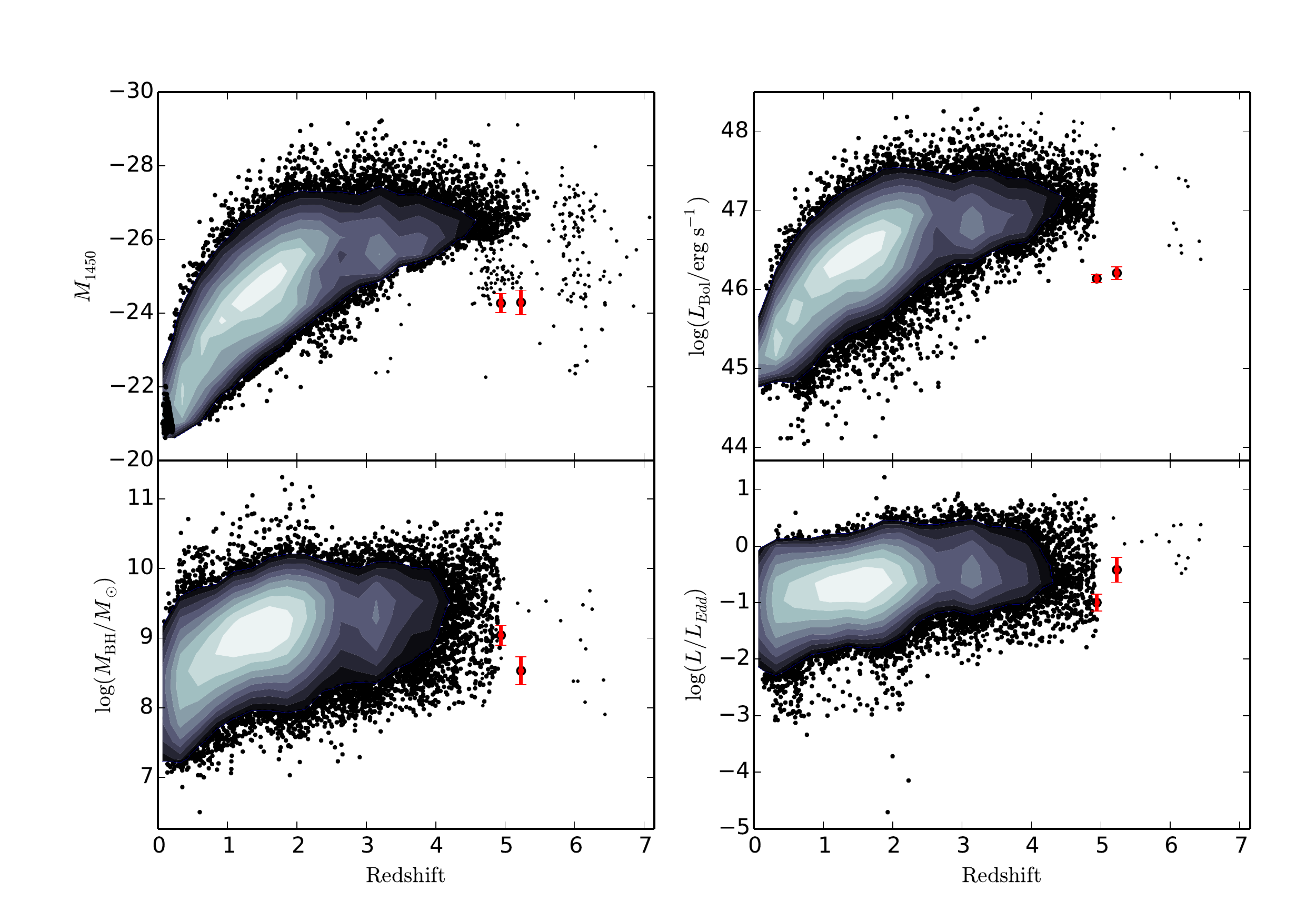}
\caption{$M_{\rm1450}$, $L_{\rm bol}$, $M_{\rm BH}$, and $L/L_{\rm Edd}$ as a function of redshift. Red and black points show our results and previous results (\citealt{2007ApJ...671.1256N,2010AJ....140..546W,2011ApJS..194...45S,2011ApJ...730....7T,2013ApJ...771...64M,2014ApJ...795L..29Y,2015ApJ...806..109J,2015Natur.518..512W,2016ApJ...828...26M,2016ApJ...825....4T,2016ApJ...819...24W}), respectively.  } 
\end{center}
\end{figure*}

\begin{figure*}[!t]
\begin{center}
\includegraphics[bb= 0 0 580 438,clip,width=18cm]{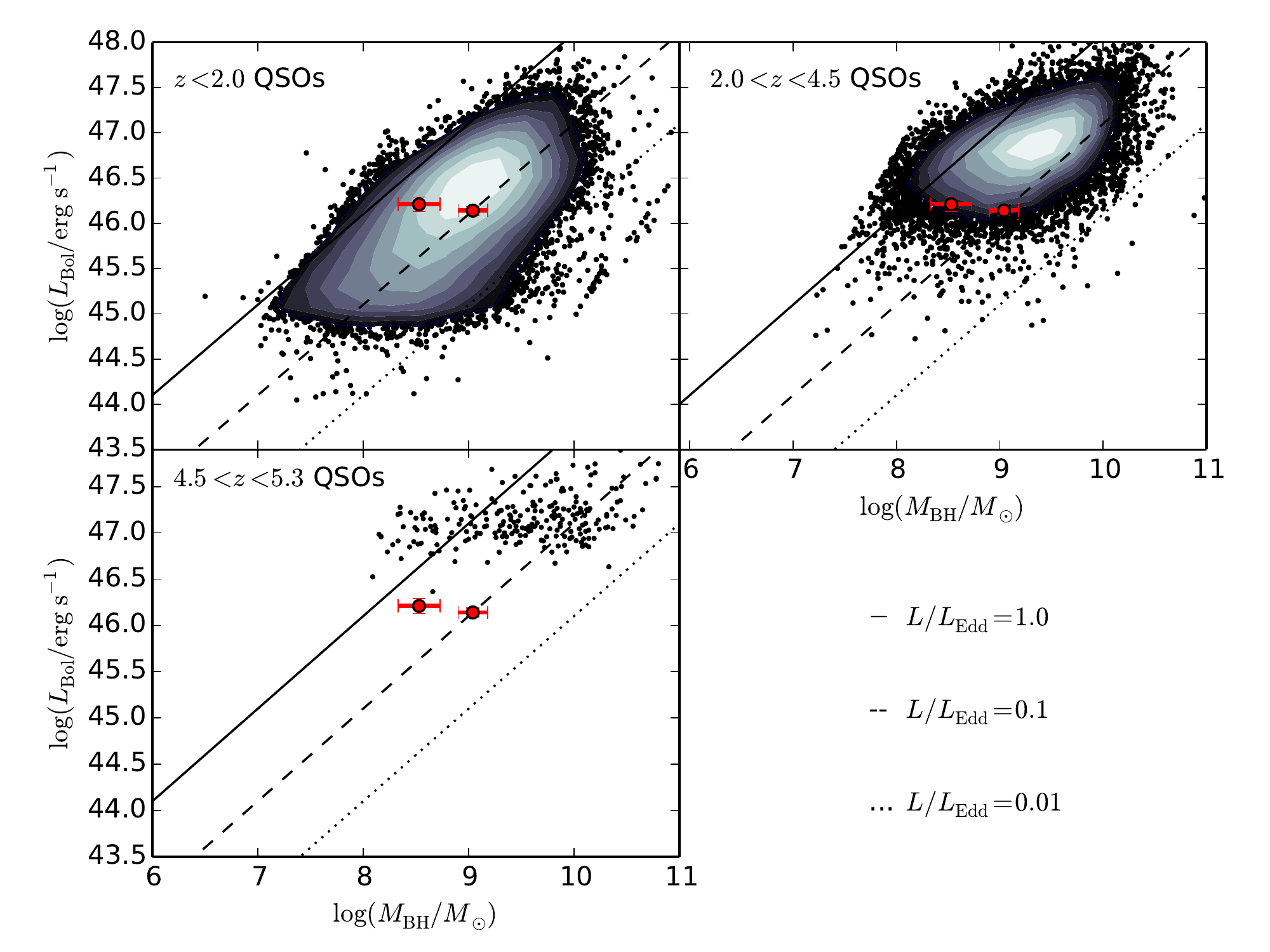}
\caption{Comparison with \Civ \ based $M_{\rm BH}$ for the SDSS quasars. Upper left and right panel, and bottom panel show $L_{\rm bol}$ vs. $M_{\rm BH}$ at $z<2.0$, $2.0<z<4.5$, and $4.5<z<5.3$, respectively. Red and black points show our results and the results of the \Civ \ based $M_{\rm BH}$ for the SDSS quasars (\citealt{2011ApJS..194...45S}), respectively. Solid, dashed, and dotted lines show the parameter space with $L/L_{\rm Edd}=$ 1.0, 0.1, and 0.01, respectively.}  
\end{center}
\end{figure*}

\begin{table*}[!hbt]

\caption{Properties of faint quasars at $z\sim5$}
\begin{center}
\begin{tabular}{@{\hspace{0.1cm}}c@{\hspace{0.1cm}}c@{\hspace{0.1cm}}c@{\hspace{0.1cm}}c@{\hspace{0.1cm}}c@{\hspace{0.1cm}}c@{\hspace{0.1cm}}c@{\hspace{0.1cm}}c@{\hspace{0.1cm}}c@{\hspace{0.1cm}}c@{\hspace{0.1cm}}c@{\hspace{0.1cm}}c@{\hspace{0.1cm}}c@{\hspace{0.1cm}}c} \hline\hline
       ID &   $M_{\rm1450}$& logFWHM$^a$$_{\rm broad}$ &logFWHM$^b$$_{\rm narrow}$&logFWHM$^c$$_{\rm tot}$& logFWHM$^d$$_{\rm used}$ &$\rm log \it M_{\rm BH}$ &$\rm log\it L_{\rm bol}$&$\rm log \it(L/L_{\rm Edd})$ &     \\
        &  &  ($\rm km~ s^{-1}$) &  ($\rm km~ s^{-1}$)&  ($\rm km~ s^{-1}$)&  ($\rm km~ s^{-1}$)&$(M_{\odot})$&$(\rm erg ~s^{-1})$&  \\ \hline

J221141.01+001118.92 & $-24.29\pm0.33$ &$3.90\pm0.12$&$3.34\pm0.02$  &$3.50\pm0.02$ &$3.50\pm0.02$& $8.53\pm0.20$ & $46.21\pm0.08$ & $-0.42\pm0.22$ &\\ 
J221520.22-000908.39& $-24.25\pm0.28$ & --&-- & -- & -- & --  &--&--&\\ 
J222216.02-000405.66 & $-24.27\pm0.26$ & $3.78\pm0.01$ &$2.98\pm0.02$ &$3.34\pm0.02$ & $3.78\pm0.01$& $9.04\pm0.14$ & $46.14\pm0.05$ & $-1.00\pm0.15$  &\\ 
\hline
              \end{tabular}              
 \end{center}
        $^{a}$ FWHM of the \Civ\ emission line for the broad component.\\
         $^{b}$ FWHM of the \Civ\ emission line for the narrow component.\\
        $^{c}$ FWHM of the \Civ\ emission line which is calculated from the derived double-Gaussian profile..\\
         $^{d}$ FWHM of the \Civ\ emission line which we used to calculate the black hole mass.\\ \\
        
\end{table*}

\section{Spectroscopic Follow-up Observations and data reduction}
We performed the spectroscopic follow-up observations of $z\sim5$ optically faint quasar candidates at the Gemini-North Telescope with the Gemini Multi-Object Spectrograph \citep[GMOS;][]{2004PASP..116..425H} on 9--10 September 2013 (HST). 
We used the R400 grating with the RG610 filter, whose wavelength coverage is 
$\rm$ 6000\AA  \ $\le$ $\rm \lambda_{obs}$  $\le$ $\rm$10000\AA. 
We used a $1 \farcs 0$-slit width, resulting in a wavelength resolution of 
$R\sim1000$ ($\Delta v$ $\sim300$ \rm km $\rm s^{-1}$). This is enough for our purposes, because the typical velocity width of quasar emission lines is wider than 2000 $\rm km$ $\rm s^{-1}$.
The typical seeing size was $\sim0 \farcs 8$.
Due to the limited observing time, 
we observed five brighter objects among nine candidates.
The individual exposure time was 900 sec, and the total exposure time was 2700 -- 3600 sec for each object (Table 1).

Standard data reduction procedures were performed by utilizing Gemini IRAF. 
After the sky subtraction,
we extracted one-dimensional spectra with an aperture size of $1 \farcs 2$. The relative sensitivity calibration was performed using the spectral data of a spectrophotometric standard star, EG131. 
The spectra of five objects were then flux-calibrated utilizing the sensitivity function which is obtained by EG131. As the \Civ \ emission line in our sample is partly absorbed by the atmospheric absorption (see Figure 3), the quasar spectra at $z\sim5$ are corrected for the atmospheric absorption by utilizing the observed spectrum of a standard star (EG131) before performing the spectral-line fitting.
Where we created the atmospheric absorption features by subtracting an artificial spectrum (which is created by IRAF task, mkspec and we assume the temperature of the black body is 11,800 K) from the obtained spectrum of EG131.
\section{Results}
We spectroscopically confirmed that three $z\sim5$ faint quasars and one $z\sim4$ quasar.
The remaining one object was identified as a late-type star. The results of our spectroscopic observations are summarized in Table 1.
Photometric and spectroscopic properties of these four faint quasars are outlined in Section 4.1.
\subsection{Notes on Individual Objects}
We summarized properties of individual objects in this section.
We note that the spectroscopic redshift of three $z\sim5$ quasars is estimated from the peak of  {C {\sc iv} $\lambda1549$} while the spectroscopic redshift of a $z\sim4$ quasar is estimated from the peak of  Ly$\alpha$.

{\itshape\bfseries J221141.01+001118.92}.
The redshift and $M_{\rm 1450}$ of this object are $z\sim5.23$ and $-24.29$, respectively.
 This object shows Ly$\alpha$ $\lambda1216$, {N {\sc v} $\lambda1240$}, {O {\sc i} $\lambda1304$}, {Si {\sc iv} $\lambda1400$}, and {C {\sc iv} $\lambda1549$.
As described in Sections 4.3 and 4.4, the inferred $\rm log \it M_{\rm BH}$  and $\rm log \it(L/L_{\rm Edd})$ are $8.53\pm0.20$ and $-0.42\pm0.22$, respectively.
\cite{2013ApJ...768..105M} also identified this object without C {\sc iv} emission lines, due to the wavelength coverage.
The estimated redshift is consistent to that of \cite{2013ApJ...768..105M}.

{\itshape\bfseries J221520.22-000908.39}.
The redshift and $M_{\rm 1450}$ of this object are $5.28$ and $-24.25$, respectively.
This object is the newly discovered faint quasars and the faintest quasars in our sample. 
Ly$\alpha$ $\lambda1216$, {N {\sc v} $\lambda1240$}, and {C {\sc iv} $\lambda1549$ emission lines are detected.
This object shows absorption lines of Ly$\alpha$ and {C {\sc iv} $\lambda1549$.
We do not estimate the black hole mass and Eddington ratio because of the absorption line of {C {\sc iv} $\lambda1549$.

{\itshape\bfseries J221941.90+001256.20}.
The redshift of this object is $4.29$. 
\cite{2013ApJ...768..105M} also identified this object.
The estimated redshift is slightly lower than that of \cite{2013ApJ...768..105M}.
A large number of absorption lines are present in the spectrum of this object, suggesting that this object could be one of the FeLoBAL quasar.
Furthermore, this object is detected in the radio wavelength (\citealt{1995ApJ...450..559B,2011AJ....142....3H}).
The peak flux density at 1.4 GHz from the Faint Images of the Radio Sky at Twenty-Centimeters (FIRST) and Very Large Array imaging of Stripe 82 is $0.87\pm 0.10$ mJy $\rm beam^{-1}$ and $0.92\pm 0.07$ mJy $\rm beam^{-1}$, respectively.
This object also has a mid-infrared ($3.4\mu \rm m$ and $4.6\mu \rm m$ with $\rm S/N>5$) counterpart in the Wide-field Infrared Survey Explorer (WISE; \citealt{2010AJ....140.1868W}), with the magnitude are 19.22  at $3.4\mu \rm m$ and 19.45 at $4.6\mu \rm m$, respectively (\citealt{2014yCat.2328....0C}).
 
{\itshape\bfseries J222216.02-000405.66}.
The redshift and $M_{\rm 1450}$ of this object are $4.94$ and $-24.27$, respectively.
Ly$\alpha$ $\lambda1216$, {N {\sc v} $\lambda1240$}, and {C {\sc iv} $\lambda1549$} emission lines are clearly detected.
As described in Sections 4.3 and 4.4, the inferred $\rm log \it M_{\rm BH}$  and $\rm log \it(L/L_{\rm Edd})$ are $9.04\pm0.14$ and $-1.00\pm0.15$, respectively.
\cite{2013ApJ...768..105M} also identified this object without C {\sc iv} emission lines, due to the wavelength coverage.
The estimated redshift is slightly lower than that of \cite{2013ApJ...768..105M}.
\subsection{Spectral-line Fitting}
In order to estimate the black hole mass, the continuum flux at 1350\AA \ and the full width at half maximum (FWHM) of the \Civ\ emission lines are needed. Since two faint quasar spectra (J221141.01+001118.92 and J222216.02-000405.66) show \Civ \ emission line without suffering from a heavy absorption line, we fit the \Civ\ emission lines of the two faint quasars.
Our fitting of the continuum flux at 1350\AA \ and the \Civ \ emission line is performed in a similar method to the one that was adopted by \cite{2013ApJ...771...64M}. The fitted wavelength range is  $\lambda_{\rm rest} \sim$ $1500-1600$\AA. Since the \Civ \ emission line in our sample seems to be asymmetry, we adopt double-Gaussian for fitting the observed {C {\sc iv}} profile (see Figure 4). As for J221141.01+001118.92 (hereafter J2211+0011), FWHM of the \Civ\ emission line is calculated from the derived double-Gaussian profile. On the other hand, as for J222216.02-000405.66 (hereafter J2222-0004),  FWHM of the \Civ\ emission line is calculated by the broad component of the {C {\sc iv}} emission line because the narrow component of the {C {\sc iv}} emission line is too narrow ($\lesssim1000$ km $\rm s^{-1}$) to explain that this comes from the broad line region.  

\subsection{Calculation of the Absolute Magnitude}
The absolute AB magnitude at 1450\AA \  of quasars is often calculated by the following equation
  (e.g., \citealt{2006AJ....131.2766R}; \citealt{2009MNRAS.399.1755C}; \citealt{2010ApJ...710.1498G}; \citealt{2012ApJ...756..160I}):
\begin{eqnarray} 
M_{1450} = m_{i'}+5-5{\rm log}d_L(z)\nonumber +2.5(1-\alpha_{\nu}){\rm log}(1+z)\\
	+2.5 \alpha_{\nu} {\rm log} \left( \frac{\lambda_{i^{\prime}}}{1450 \AA} \right), 
\end{eqnarray} 
where $d_L(z)$, $\alpha_{\nu}$, and $\lambda_{i^{\prime}}$ are the luminosity distance, spectral index of the quasar continuum ($f_{\nu}\propto\nu^{-\alpha_{\nu}}$, where the typical $\alpha_{\nu}=0.5$}; \citealt{2006AJ....131.2766R}), and the effective wavelength of the $i'$-band, respectively. In order to calculate $M_{\rm 1450}$ as accurate as possible, we calculate $M_{\rm 1450}$ from the median flux of the obtained spectra between rest-frame $\lambda = 1425$\AA \ and $1475$\AA. The calculated $M_{\rm 1450}$ is listed in Table 2.

\subsection{Estimates of the Black Hole Masses and Eddington Ratios}
Since two faint quasars show {C {\sc iv} $\lambda1549$} with no absorption lines, we estimate the black hole mass of them by the following equation (e.g., \citealt{2011ApJS..194...45S}):
\begin{equation}
\log\left(\frac{M_{\rm BH,vir}}{M_\odot}\right)= a + b\log\left({\lambda L_{\lambda}(1350 \AA) \over 10^{44}\,{\rm
erg\,s^{-1}}}\right) + c\log\left(\frac{\rm FWHM}{\rm km\,s^{-1}}\right)\ ,
\end{equation}
where $\lambda L_{\lambda}$, $a$, $b$, and $c$ is the luminosity, 0.66, 0.53, and 2, respectively. The luminosity at 1350 \AA, $\lambda L_\lambda(1350 \rm{\AA})$, can be calculated as follows:
\begin{equation}
 \lambda L_\lambda(1350 \AA)=\lambda\it \ F_\lambda(\rm1350 \rm{\AA}) \ 4\pi\it d_L (z)^{\rm2},
\end{equation}
where $ F_\lambda(\rm1350 \AA)$ is calculated from the individual quasar spectrum.
The Eddington ratio, $L/L_{\rm Edd}$ is calculated as follows (e.g., \citealt{2013BASI...41...61S}):
\begin{equation}
L/L_{\rm Edd}  = L_{\rm bol} / \left(1.26\times10^{38} M_{\rm BH}/M_{\odot}\right),
\end{equation}
where $L_{\rm Edd}$ and $L_{\rm bol}$ are the Eddington luminosity and the bolometric luminosity, respectively. 
$L_{\rm bol}$ can be calculated as follows:
\begin{equation}
\label{eq_bol}
L_{\rm bol}=f_{\rm BC} \ \lambda L_\lambda(1350 \rm{\AA}),
\end{equation}
where $f_{\rm BC}$ is the bolometric correction factor. In this study, we use $f_{\rm BC}=3.81 $ (\citealt{2006ApJS..166..470R}).
Using equations (15)--(18), we estimate the black hole mass and Eddington ratio. The estimated black hole mass and Eddington ratio are listed in Table 2.

\begin{figure*}[!t]
\begin{center}
\includegraphics[bb= 40 0 770 698,clip,width=17cm]{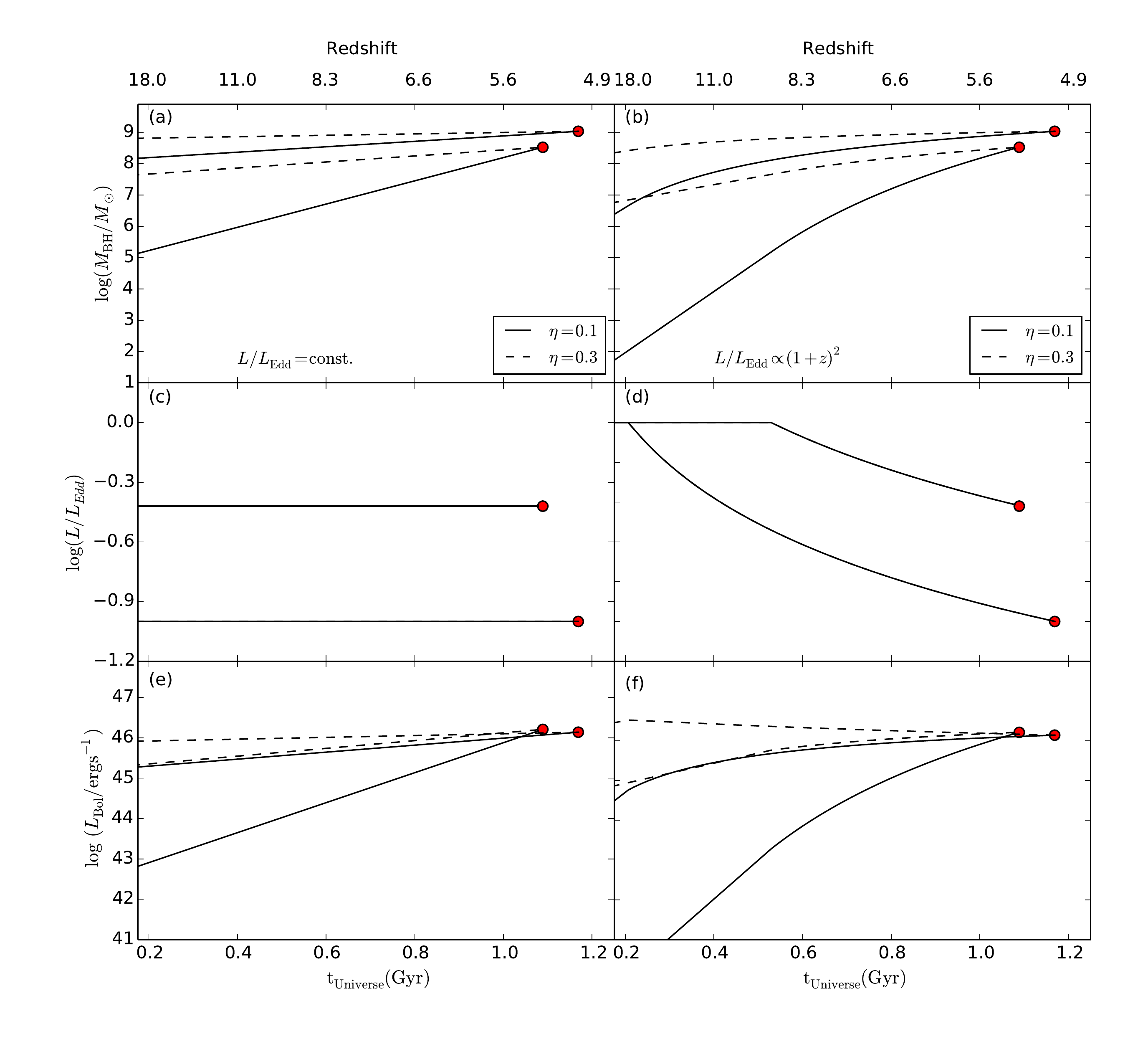}
\caption{ Evolutionary tracks of the SMBHs in our sample. Upper left and right panels show $M_{\rm BH}$ vs. $t_{\rm Universe}$. Second left and right panels show $\rm log(\it L/L_{\rm Edd}) $ vs. $t_{\rm Universe}$. Third left and right panel show $L_{\rm Bol}$ vs. $t_{\rm Universe}$. Solid and dashed lines show the evolutionary tracks for $\eta=0.1$ and $\eta=0.3$, respectively. All left and right panels are assumed that $\it L/L_{\rm Edd}=$ constant and $L/L_{\rm Edd}\propto(1+z)^{2}$, respectively.}  
\end{center}
\end{figure*}

\begin{figure*}[!t]
\begin{center}
\includegraphics[bb= 30 0 780 695,clip,width=17cm]{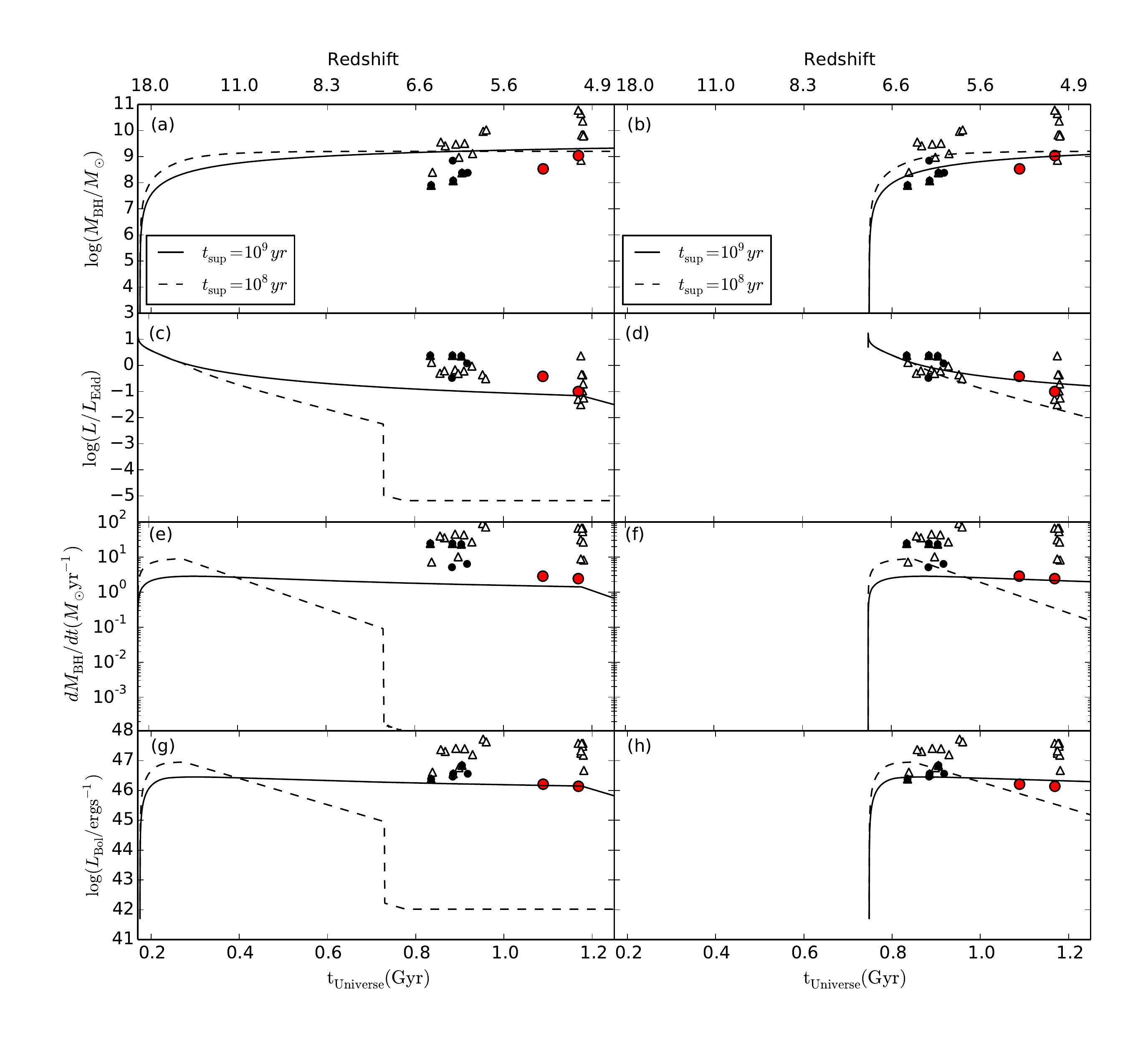}
\caption{Comparison of observational results and theoretical results. Upper left and right panels show $\rm log (\it M_{\rm BH}/M_{\odot})$ vs. $t_{\rm universe}$. Second left and right panels show $\rm log(\it L/L_{\rm Edd}) $ vs. $t_{\rm universe}$. Third left and right panels show $dM_{\rm BH}/dt$ vs. $t_{\rm universe}$. Fourth left and right panel show $L_{\rm Bol}$ vs. $t_{\rm universe}$. Red, black circle, and triangles show our results at $z\sim5$, faint quasars at $z\sim6$ (\citealt{2010AJ....140..546W}), and luminous quasars at $\sim5-6$ (\citealt{2007AJ....134.1150J,2010AJ....140..546W,2011ApJS..194...45S}), respectively. Solid and dashed lines show the evolutionary tracks for $t_{\rm sup}=10^9~\rm yr$ and $t_{\rm sup}=10^8~\rm yr$, respectively. All left and right panels assume that the seed BHs of the $z\sim5$ faint quasars are formed at $z\sim20$ and $z\sim7$, respectively.}  
\end{center}
\end{figure*}


\section{Discussion}
\subsection{Comparison with the previous results}
As reported in Section 3, we have selected nine faint quasar candidates at $z\sim5$.
Then we have performed the spectroscopic observation of five objects among the quasar candidates at $z\sim5$ and
found that four objects are faint quasars.
As we mentioned in Section 2.3, several studies reported that quasar candidates are selected by utilizing not only optical data but also the near-infrared data (e.g., \citealt{2011AJ....142...78W,2013ApJ...768..105M}).
To examine whether it is really useful for selecting quasars by adding the near-infrared data, we check the near-infrared colors of the spectroscopically confirmed objects.
As shown in Figure 2, all of the spectroscopically confirmed faint quasars are selected and one spectroscopically confirmed star is removed by adding the near-infrared selection criteria for quasars at $z\sim5$. 
Thus, we conclude that it is useful to distinguish contaminants and faint quasars by adding the near-IR data.   

Since our survey field is overlapped with the SDSS stripe 82 field and the luminosity range of $z\sim5$ faint quasars which we identified is almost the same, we just check the expected number of faint quasars. If we assuming that the completeness is unity, the expected number of faint quasars is calculated by the following equation:
\begin {equation}\label{phist}
N\sim\int_{m_{i}=21}^{m_{i}=23}  \Phi(m_{i}, z\sim5)~ dm_{i} V_{\rm c}(z),
\end{equation} 
where  $\Phi(m_{i}, z\sim5)$ and  $V_{\rm c}(z)$ are the $z\sim5 $ quasar luminosity function which is derived by \cite{2013ApJ...768..105M} and the comoving volume in our survey at $4.7<z<5.3$ ($V_{\rm c}(z)\sim3.5\times10^7$Mpc${^3}$), respectively.
The expected number of $z\sim5$ faint quasars in our survey, $N$ is $\sim3.6$. On the other hand, we select nine candidates by utilizing the optical data and we then select five quasar candidates among them by adding the near-infrared data. Four quasar candidates among five objects have been carried out the spectroscopic follow-up observations and three objects and one object are identified as $z\sim5$ faint quasars and $z\sim4$ faint quasar, respectively. Therefore the success rate is 0.75. Since there is one object which we have not yet performed the spectroscopic observation, the corrected number of $z\sim5$ faint quasars in this survey is 3.75. This result is roughly consistent with the expected number of $z\sim5$ faint quasars in our survey. This suggests that the completeness is not so low even if we add the near-infrared data to select $z\sim5$ faint quasars effectively. 

We plot our faint $z\sim5$ quasars in the redshift-$M_{\rm 1450}$ space, the redshift-$L_{\rm bol}$ space, the redshift-$M_{\rm BH}$ space, and the redshift-$\it(L/L_{\rm Edd})$ space, respectively (Figure 5). As shown in Figure 5, our quasar sample is relatively faint among known quasar samples (e.g., \citealt{2011ApJS..194...45S}) and the black hole masses and Eddington ratios of our sample are relatively lower than those of the known quasar sample. However, the masses of SMBHs are estimated by the various emission lines and equations (e.g., \citealt{2004ApJ...614..547S, 2007AJ....134.1150J, 2007ApJ...669...32K, 2007ApJ...671.1256N, 2010AJ....140..546W,2011ApJ...742...93A,2011ApJ...739...56D,2011ApJS..194...45S,2011ApJ...730....7T,2012ApJ...761..143N,2012ApJ...753..125S,2013ApJ...771...64M,2013ApJ...770...87P,2014ApJ...790..145D,2014ApJ...795L..29Y,2015ApJ...806..109J,2015ApJ...815..128K,2015Natur.518..512W,2016PASJ...68...40M,2016PASJ...68....1S,2016ApJ...825....4T}) and some of them are plotted in Figure 5. They depend on which emission lines and equations are used to estimate the black hole mass. Therefore it is difficult to compare the black hole mass of our sample with that of previous studies if the black hole mass is calculated by the different emission line and the different equation. Thus, we have to compare the black hole mass of our sample with that of previous studies which was calculated by the same emission line and the same equation. 

Since the black hole mass and Eddington ratio of our sample are calculated in the same manner as described by \cite{2011ApJS..194...45S}, we compare our sample with \Civ~ based $M_{\rm BH}$ for the SDSS quasars (Figure 6). The median of $\rm log (\it M_{\rm BH}/M_{\odot})$ and $\rm log(\it L/L_{\rm Edd}) $ for the SDSS quasars at $4.5<z<5.3$ are 9.03 and -0.53, respectively. Therefore $M_{\rm BH}$ of J2211+0011 is lower than that of luminous SDSS quasars at similar redshift. On the other hand, $M_{\rm BH}$ of J2222-0004 is similar to that of  luminous SDSS quasars at similar redshift. As for the Eddington ratio, the Eddington ratio of J2211+0011 is higher than that of luminous SDSS quasars at similar redshift. On the other hand, the Eddington ratio of J2222-0004 is lower than that of luminous SDSS quasars at similar redshift. 
\subsection{Constraints on the growth history of SMBHs}

\begin{table*}[!hbt]
\begin{center}
\caption{Summary of the expected $M_{\rm seed}$ for our $z\sim5$ faint quasars}
\begin{tabular}{ccc@{\hspace{0.5cm}}c@{\hspace{0.5cm}}c@{\hspace{0.5cm}}c@{\hspace{0.5cm}}c@{\hspace{0.5cm}}c@{\hspace{0.5cm}}c@{\hspace{0.1cm}}c@{\hspace{0.1cm}}c@{\hspace{0.1cm}}c@{\hspace{0.1cm}}c@{\hspace{0.1cm}}c@{\hspace{0.1cm}}c@{\hspace{0.1cm}}c@{\hspace{0.1cm}}c@{\hspace{0.1cm}}c@{\hspace{0.1cm}}c@{\hspace{0.01cm}}c} \hline\hline
   $L/L_{\rm Edd}$    &$\eta$ & $z_{\rm seed}^{a}$&  $M_{\rm seed}$ of J2211+0011 & $M_{\rm seed}$ of J2222-0004 &    \\ \hline

const.&0.1 & 20 &$\sim10^{5}M_{\odot}$ &  $\sim10^{8}M_{\odot}$ & \\ 
const. &0.3 & 20&$\sim10^{7-8}M_{\odot}$ &  $\sim10^{8}M_{\odot}$ & \\ 
$\propto(1+z)^2$ &0.1 & 20&$\sim10^{1-2}M_{\odot}$ &  $\sim10^{6}M_{\odot}$ & \\ 
$\propto(1+z)^2$ &0.3 & 20&$\sim10^{7}M_{\odot}$ &  $\sim10^{8}M_{\odot}$ & \\ 
model A$^{b}$ &0.1 & 20&$\sim10^{3}M_{\odot}$ &  $\sim10^{3}M_{\odot}$ & \\ 
model B$^{c}$ &0.1 & 7&$\sim10^{3}M_{\odot}$ &  $\sim10^{3}M_{\odot}$ & \\ 

\hline
              \end{tabular}              
      \end{center}
         $^{a}$ The redshift which the seed black holes formed.\\
         $^{b}$ The super-Eddington growth model of \cite{2008ApJ...681...73K} with $z_{\rm seed}\sim20$.\\
         $^{c}$ The super-Eddington growth model of \cite{2008ApJ...681...73K} with $z_{\rm seed}\sim7$.\\
         \end{table*}

\subsubsection{Estimates the growth time and evolutionary track of SMBHs}
Since we calculate $M_{\rm BH}$ and $ L/L_{\rm Edd}$ in Section 4.4, we can investigate the growth time of the SMBHs which is one of the important parameter to constrain the growth history of SMBHs in our quasar sample.
To investigate the growth time of the SMBHs in our sample, we calculate the growth time, $ t_{\rm growth}$ as follows:
\begin{equation}
  t_{\rm growth} = \tau ~ \ln
  \left( \frac{M_{\rm BH}}{M_{\rm seed}} \right) (1/f_{\rm active}) ~ {\rm yr},
\end{equation}
where $M_{\rm seed}$ is the seed black hole mass. $\tau$ is given as follows:
\begin{equation}
\tau = 4.5 \times 10^8 ~~ \frac {\eta /(1- \eta) }{L/L_{\rm Edd}} ~ {\rm yr}, 
\end{equation}
where $\eta$ is the radiative efficiency, and $\eta=0.054$ and $0.42$ correspond to the non-rotating Schwarzschild BH (\citealt{1916MathPhys..XX..189}) and the maximally rotating Kerr BH (\citealt{1963PhRvL..11..237K}), respectively. Moreover $f_{\rm active}$ is the duty cycle (the fraction of the active time, \citealt{2007ApJ...671.1256N}). There are various scenarios for the formation of seed black holes (see \citealt{2010A&ARv..18..279V}). One of such scenarios assumes that the seed BHs are the remnants of Population III stars (e.g., \citealt{2001ApJ...551L..27M}). In this case, $M_{\rm seed}\sim10-100M_{\odot}$. Another important scenario is that the seed BHs have been formed by direct collapse model (e.g., \citealt{1994ApJ...432...52L}). In this case, $M_{\rm seed}\sim10^5M_{\odot}$. As stated above, there are various range of $M_{\rm seed}$ and it is also important to constrain $M_{\rm seed}$ of $z\sim5$ faint quasars. 

From this kind of circumstances, we first assume $\eta=0.1$, $f_{\rm active}=1$, and $M_{\rm seed}=10M_{\odot}$ at $z\sim20$ as a very simple assumption. In this case, $t_{\rm growth}$ is $\sim1\rm Gyr$ even if we assume that $ L/L_{\rm Edd}$ is 1 almost up to $z\sim6$ and then dropped to the observed values of 0.1 (J2222-0004) or 0.4 (J2211+0011) at $z\sim5$. Since $ L/L_{\rm Edd}$ of our sample is $\sim0.1-0.4$, $M_{\rm 1450}$ of our sample could be $\sim-26$ to $-27$ at $z\sim6$ from Equation (17).
Therefore if $z\sim6$ quasars at $M_{\rm 1450}\sim -26$ to $-27$ and our $z\sim5$ faint quasars are the same populations (i.e., the progenitors of our $z\sim5$ faint quasars are $z\sim6$ luminous quasars), the $z\sim6$ number count at this magnitude range is expected to be $\sim10^{-7}$ $\rm Mpc^{-3}$ $\rm mag^{-1}$ because the number count of our quasar sample at $z\sim5$ is $\sim10^{-7}$ $\rm Mpc^{-3}$ $\rm mag^{-1}$ (\citealt{2013ApJ...768..105M}). On the other hand, it is reported that the $z\sim6$ number count at $M_{1450}\sim-26$ to $-27$ is $\sim10^{-9}$ $\rm Mpc^{-3}$ $\rm mag^{-1}$ from the quasar survey (\citealt{2010AJ....139..906W}). In order to explain without contradiction with QLF studies, it is required that $z\sim6$ quasars at $M_{\rm 1450}\sim -26$ to $-27$ and our $z\sim5$ faint quasars are not the same populations but different populations (i.e., the progenitors of our $z\sim5$ faint quasars are not $z\sim6$ luminous quasars). Moreover, it is difficult to explain our growth with $M_{\rm seed}=10M_{\odot}$ at $z_{\rm seed}$ even under this most rapid Eddington-limited growth scenario to explain without contradiction with QLF studies. It is therefore suggested that $M_{\rm seed}$ of our $z\sim5$ faint quasars is greater than $10M_{\odot}$.

In order to compare with previous results (\citealt{2007ApJ...671.1256N,2011ApJ...730....7T}), we assume $M_{\rm seed}=10^{4} M_{\odot}$ with the same $\eta$ and $f_{\rm active}$. In addition, we use $L/L_{\rm Edd}$ ($\rm log \it(L/L_{\rm Edd})=\rm -1.00$ and $-0.42$ for J2222-0004 and J2221+0011, respectively.), which is calculated by Equation (17). The calculated $t_{\rm growth}$ are 5.80 and 1.37 Gyr for J2222-0004 and J2221+0011, respectively.
We then calculate $ t_{\rm growth}/ (t_{z\rm QSO} - t_{z\rm seed})$, where $t_{z\rm QSO}$ and $t_{z\rm seed}$ are the time of the source (i.e., at the redshift of the source) and the formation time of the seed black hole, respectively. The calculated $ t_{\rm growth}/ (t_{z\rm QSO} - t_{z\rm seed})$ of J2222-0004 and J2211+0011 are 5.84 and 1.50, respectively. Therefore it is expected that J2222-0004 and J2211+0011 has experienced in growing phase with higher $L/L_{\rm Edd}$ in the past because $ t_{\rm growth}/  (t_{z\rm QSO} - t_{z\rm seed}) >1$. 

We also estimate the evolutionary tracks of the SMBHs in our sample up to $z\sim20$ (Figure 7) because the value of $M_{\rm seed}$ is important parameter to investigate the growth history of SMBHs.
Since $\eta$ is poorly constrained (the range of $\eta$ is $\sim0.05-0.3$), we assume that $\eta$ is 0.1 firstly and then we estimate $M_{\rm seed}$ of $z\sim5$ faint quasars at $z\sim20$.
We also assume that $ L/L_{\rm Edd}$ $=$ constant up to $z\sim20$ as a simple assumption at first.
As shown in the left side of Figure 7, it is expected that $M_{\rm seed}$ of J2222-0004 and J2211+0011 are $\sim10^{8}M_{\odot}$ and $\sim10^{5}M_{\odot}$, respectively. Next, we assume that  $\eta=0.3$ as a most difficult case to grow the SMBHs. Then we estimate $M_{\rm seed}$ at $z\sim20$.
As a result, it is expected that $M_{\rm seed}$ of J2222-0004 and J2211+0011 are $>10^{8}M_{\odot}$ and $\sim10^{7-8}M_{\odot}$, respectively.  This result suggests that $M_{\rm seed}$ of $z\sim5$ faint quasars in our sample are needed to be massive black holes ($>10^{7}M_{\odot}$). These results are summarized in Table 3. As shown in (e) of Figure 7, the luminosity of our $z\sim5$ faint quasars is expected to be higher with increasing $t_{\rm univ}$. Therefore it is expected that evolutionary track of $z\sim6$ luminous quasars and $z\sim5$ faint quasars are not the same. Thus, they are not the same populations but different populations in this case.

While we assumed that $L/L_{\rm Edd}$ = constant, it may also well vary with time.
In fact, it is reported that $L/L_{\rm Edd}\propto(1+z)^{2}$ is consistent with a fit to the observational data at $2\lesssim z\lesssim 6.5$ (\citealt{2016ApJ...825....4T}). Therefore we consider a scenario where $L/L_{\rm Edd}$ increases with $(1+z)^{2}$ until $L/L_{\rm Edd}=1$ as a more realistic assumption.
In the case of $\eta=0.1$, it is expected that $M_{\rm seed}$ of J2222-0004 and J2211+0011
 are $\sim10^{6}M_{\odot}$ and $\sim10^{1-2}M_{\odot}$, respectively.
In the case of $\eta=0.3$, it is expected that $M_{\rm seed}$ of J2222-0004 and 
J2211+0011 are $\sim10^{8}M_{\odot}$ and $\sim10^{7}M_{\odot}$, respectively. These results are also summarized in Table 3.
As shown in (f) of Figure 7, the luminosity of our $z\sim5$ faint quasars is expected to be higher with increasing $t_{\rm univ}$ or similar luminosity. Therefore the evolutionary track of $z\sim6$ luminous quasars and $z\sim5$ faint quasars are not the same, suggesting that they are not the same populations but different populations in this case also.

From these results, it can be concluded that $M_{\rm seed}$ of $z\sim5$ faint quasars in our sample are expected to be $>10^{5}M_{\odot}$ in most cases if we assume that $ L/L_{\rm Edd}$$=$constant or $L/L_{\rm Edd}\propto(1+z)^{2}$. On the other hand, previous study reported that a median value of $M_{\rm BH}$ and $L/L_{\rm Edd} $ for the SDSS luminous quasars at $z\sim4.8$ are $\sim8.4\times 10^8M_{\odot}$ ($10^8M_{\odot}\lesssim M_{\odot} \lesssim 6.6 \times 10^9 M_{\odot}$) and $\sim0.6$ ($0.2 \lesssim L/L_{\rm Edd}\lesssim 3.9$), respectively (\citealt{2011ApJ...730....7T}). They also reported that $\sim40\%$ of the SDSS luminous quasars at similar redshift could have been formed at $M_{\rm BH}<10^2M_{\odot}$, and  
$L/L_{\rm Edd}$ of the $z\sim5$ faint quasars in our sample is relatively lower than that of the most of luminous quasars at $z\sim5$. This is the main reason that $M_{\rm seed}$ of $z\sim5$ quasars in our sample is much larger than that of them. Therefore these results may be suggesting that $M_{\rm seed}$ of $z\sim5$ quasars depends on the luminosity. In addition, $z\sim6$ luminous quasars and $z\sim5$ faint quasars are not the same populations but different populations in all cases.

We note that many previous studies mentioned that the black hole mass which is estimated by the \Civ\ emission line has large uncertainty (e.g., \citealt{2012MNRAS.427.3081T,2012ApJ...753..125S}).
In order to do the more accurate statistical discussions, we have to construct the larger samples of faint quasars at high redshift and we need to use not only \Civ\ but also other emission lines.
\subsubsection{Comparison with the theoretical model}
As we have discussed in Section 5.2.1, we have assumed that $L/L_{\rm Edd}$ is constant or $L/L_{\rm Edd}\propto(1+z)^{2}$.
However, the influence of a mass outflow, which is the strong radiation pressure from the accretion disk (e.g., \citealt{2005ApJ...628..368O,2007ApJ...659..205O}), is not considered in these assumptions. The mass accretion history of SMBHs is affected by this mass outflow. Therefore, the assumed mass accretion history of SMBHs is physically unrealistic.
In order to investigate the
seed black hole mass of $z\sim5$ faint quasars by a more realistic mass accretion history of the SMBHs,
we compare with the theoretical model of \cite{2008ApJ...681...73K} (see also \citealt{2009ApJ...706..676K}).
In this model,  the mass accretion rate onto a central BH is driven by the 
turbulent viscosity due to supernova (SN) explosions. Moreover, the gas supply rate of 
the circumnuclear disk (CND) from the host galaxies regulates the maximal black hole accretion rate. 
There are two types of models: Eddington-limited growth models and super-Eddington growth models.
Both growth models include the influence of a mass outflow due to the strong radiation pressure from the accretion disk (e.g., \citealt{2005ApJ...628..368O,2007ApJ...659..205O}) and they assume that the mass of the seed BHs is $10^3$ $M_{\odot}$. According to the radiation hydrodynamic simulations, the super-Eddington accretion, could be possible (e.g., \citealt{2005ApJ...628..368O,2007ApJ...659..205O}). Therefore, we use the super-Eddington growth model of \cite{2008ApJ...681...73K} to compare it with the observational data.

Since the formation time of the seed BHs is another important parameter to constrain the growth history of the SMBHs, we here examine the two scenarios of $z\sim5$ faint quasar formation and evolution (Figure 8). The first scenario is that the case of the seed BHs of our $z\sim5$ faint quasars formed at $z\sim20$ (left panels of Figure 8). In this case, $M_{\rm BH}$ of our $z\sim5$ faint quasars can be reproduced even if $M_{\rm BH}$ of the seed BHs is $\sim10^3M_{\odot}$. In addition, $dM_{\rm BH}/dt$ of our $z\sim5$ quasars and $z\sim5-6$ quasars from the literature (e.g., \citealt{2010AJ....140..546W,2011ApJS..194...45S}) are roughly consistent with the case of a period of the 
mass supply from host galaxies, $t_{\rm sup} =10^9$ yr rather than the case of $t_{\rm sup}=10^8$ yr. The time between the peak and rapid decline of the AGN luminosity corresponds to the quasar phase in this model (see also Figures 6 and 7 of \citealt{2008ApJ...681...73K}). 
As discussed, since the number count of our $z\sim5$ quasars is some 2 orders of magnitude larger, the quasar lifetime, $t_{\rm QSO}$ may be longer than SDSS high-$z$ quasars, if the bias values are similar. As shown in (a) and (c) of Figure 8, it seems that the evolutionary tracks of $z\sim6$ luminous quasars and our $z\sim5$ faint quasars are the same. However, the $z\sim6$ and $z\sim5$ quasars have the similar luminosity in this scenario (see (g) of Figure 8). Therefore it can be concluded that $z\sim6$ luminous quasars and our $z\sim5$ faint quasars are not the same populations but different populations under this scenario also.

Yet another scenario is that the case of the seed BHs of the $z\sim5$ faint quasars in our sample formed at $z\sim7$ (right panels of Figure 8).
In this case, $M_{\rm BH}$ of the $z\sim5$ faint quasars also can be reproduced even if $M_{\rm BH}$ of the seed BHs is $\sim10^{3}M_{\odot}$. Furthermore, $dM_{\rm BH}/dt$ of our $z\sim5$ faint quasars is roughly consistent with the case of $t_{\rm sup} =10^9$ yr. On the other hand, $dM_{\rm BH}/dt$ of $z\sim6$ quasars are roughly consistent with the case of $t_{\rm sup} =10^8$ and $t_{\rm sup} =10^9$ yr. In the case of $t_{\rm sup} =10^9$ yr, it seems that $z\sim6$ luminous quasars and our $z\sim5$ faint quasars are on the same evolutionary track (see (b) and (d) of Figure 8). However $z\sim6$ and $z\sim5$ quasars have the similar luminosity. Therefore it is expected that the evolutionary tracks of $z\sim6$ luminous quasars and our $z\sim5$ faint quasars are not the same. This suggests, again, that $z\sim6$ luminous quasars and our $z\sim5$ faint quasars are not the same populations but different populations.

 In conclusion, we find three main results by comparing with the super-Eddington growth model of \cite{2008ApJ...681...73K}. Firstly, we confirm that $M_{\rm BH}$ of our $z\sim5$ faint quasars can be reproduced even if $M_{\rm BH}$ of the seed BHs is $\sim10^3M_{\odot}$. Secondly, we find that the origin for the difference of $z\sim6$ luminous quasars and our $z\sim5$ faint quasars may be explained by the difference of $t_{\rm sup}$. Lastly, we confirm that it can be explained $M_{\rm BH}$ of  $z\sim6$ luminous quasars and our $z\sim5$ faint quasars even if the seed BHs of them are formed at $z\sim7$.

 To investigate whether $z\sim6$ luminous quasars and our $z\sim5$ faint quasars are the same populations or not, the information about the gas of the $z\sim5-6$ quasar host galaxies is needed. This is because $M_{\rm gas}$ (the gas mass in a circumnuclear disk, see Figure 1 of \citealt{2008ApJ...681...73K}) $ / M_{\rm BH}$ of $z\sim5$ quasars is expected to be larger than that of $z\sim6$ quasars (see Figure 8 in \citealt{2009ApJ...706..676K}) if $z\sim6$ and $z\sim5$ quasars are not the same. This important issue can be investigated by submillimeter observations such as the Atacama Large Millimeter/submillimeter Array (ALMA).

\subsubsection{Influence by the uncertainty of the BH Mass Estimates}
As we have stated in Section 5.2.1, many previous studies reported that the black hole mass which is estimated by the \Civ\ emission line has large uncertainty (e.g., \citealt{2012MNRAS.427.3081T, 2012ApJ...753..125S}).
On the other hand, several previous studies attempted to reduce this uncertainty (e.g., \citealt{2012ApJ...759...44D,2013MNRAS.434..848R,2013ApJ...770...87P,2016MNRAS.461..647C,2017MNRAS.465.2120C,2017ApJ...839...93P}).
Among them, we especially focus on the recent result of \cite{2016MNRAS.461..647C,2017MNRAS.465.2120C}. 
\cite{2016MNRAS.461..647C} reported that the FWHM of the \Civ\ emission line is correlated with the \Civ\ blueshift (Figure 7 of \citealt{2016MNRAS.461..647C}). Furthermore, they found that the black hole masses which are derived by the \Civ\ emission line of quasars with the low \Civ\ blueshift (\Civ\ FWMH $\lesssim1000$ $\rm km$ $\rm s^{-1}$) are systematically underestimated while the black hole masses of quasars with high \Civ\ blueshift (\Civ\ blueshift $\gtrsim1000$ $\rm km$ $\rm s^{-1}$) are overestimated (see also \citealt{2017MNRAS.465.2120C}).

Since the \Civ\ FWHM of J2222-0004 is $\sim6000$ $\rm km$ $\rm s^{-1}$, it is expected that the \Civ\ blueshift of this object is $\sim1000$ $\rm km$ $\rm s^{-1}$. Thus, the corrected black hole mass and Eddington ratio of this object are not changed. As for J2211+0011, the \Civ\ FWHM of J2211+0011 is $\sim3200$ $\rm km$ $\rm s^{-1}$ and this corresponds to the \Civ\ blueshift of this object is $<1000$ $\rm km$ $\rm s^{-1}$. Therefore the black hole mass of J2211+0011 is underestimated and the corrected Eddington ratio of this object could be lower than the uncorrected Eddington ratio of this object. In this case, $M_{\rm seed}$ of J2211+0011 at $z\sim20$ becomes large if we assume that $ L/L_{\rm Edd}$$=$constant or $L/L_{\rm Edd}\propto(1+z)^{2}$. However, the main conclusion of Section 5.2.1 is not any changed even if we consider the influence by the uncertainty of the BH Mass Estimates. The conclusions of Section 5.2.2 are also not changed at all.

\section{Summary}
 We have searched for optically faint quasars at $z\sim5$ in a part of the CFHTLS wide field ($\sim6$ deg$^2$) to investigate the black hole mass, Eddington ratio, and the growth history of the SMBHs in our sample. The main results of our works are briefly summarized below.

\begin{enumerate}

\item Utilizing the CFHTLS wide and UKIDSS DXS catalog, we selected nine $z\sim5$ faint quasar candidates and we then performed spectroscopic observation of five objects among them. Then we confirmed that three $z\sim5$ faint quasars, a $z=4.29$ faint quasar, and a late-type star. We also confirmed that near-infrared data is useful to distinguish contaminants and $z\sim5$ faint quasars effectively.

\item We estimated the black hole mass and Eddington ratio of two $z\sim5$ faint quasars based on the broad \Civ\ line. The inferred $\rm log \it M_{\rm BH}$ are $9.04\pm0.14$ and $8.53\pm0.20$, respectively. In addition, the inferred $\rm log \it(L/L_{\rm Edd})$ are $-1.00\pm0.15$ and $-0.42\pm0.22$, respectively.

\item It is expected that $M_{\rm seed}$ of $z\sim5$ faint quasars in our sample are $>10^{5}M_{\odot}$ in most cases if we assume that $ L/L_{\rm Edd}$ $=$ constant or $L/L_{\rm Edd}\propto(1+z)^{2}$. This result may be suggesting that $M_{\rm seed}$ of $z\sim5$ quasars depends on the luminosity because previous study reported that $\sim40\%$ of the SDSS luminous quasars at similar redshift could have been formed at $M_{\rm BH}<10^2M_{\odot}$.

\item If we compare with the theoretical model, $M_{\rm BH}$ of the $z\sim5$ faint quasars in our sample can be reproduced even if $M_{\rm BH}$ of the seed BHs is $\sim10^{3}M_{\odot}$. 

\item Since $z\sim6$ luminous quasars and our $z\sim5$ faint quasars are not on the same evolutionary track, $z\sim6$ luminous quasars and our $z\sim5$ faint quasars are not the same populations but different populations. The origin for the difference of $z\sim6$ luminous quasars and our $z\sim5$ faint quasars may be explained by the difference of $t_{\rm sup}$. 

\item We confirm that it can be explained $M_{\rm BH}$ of  $z\sim6$ luminous quasars and our $z\sim5$ faint quasars even if the seed BHs of them are formed at $z\sim7$.
\end{enumerate}

\acknowledgments

We thank the anonymous referee for valuable comments helping to improve this paper.
We also would like to thank the Gemini staff for their invaluable helps. This work is based on data obtained at the Gemini Observatory via the time exchange program between Gemini and the Subaru Telescope (processed using the Gemini IRAF package). The Gemini Observatory is operated by the Association of Universities for Research in Astronomy, Inc., under a cooperative agreement with the NSF on behalf of the Gemini partnership: the National Science Foundation (United States), the National Research Council (Canada), CONICYT (Chile), Ministerio de Ciencia, Tecnologia e Innovaci\'{o}n Productiva (Argentina), and Minist\'{e}rio da Ci\^{e}ncia, Tecnologia e Inova\c{c}\~{a}o (Brazil).

This work is based on observations performed with MegaPrime/MegaCam, that is a joint project of CFHT and CEA/DAPNIA, at the Canada-France-Hawaii Telescope (CFHT) which is operated by the National Research Council (NRC) of Canada, the Institut National des Science de l'Univers of the Centre National de la Recherche Scientifique (CNRS) of France, and the University of Hawaii. This work is based in part on data products produced at the Canadian Astronomy Data Centre as part of the Canada-France-Hawaii Telescope Legacy Survey, a collaborative project of NRC and CNRS. To make contour in Figures 2, 5, and 6, the astroML python package is used (\citealt{astroML,astroMLText}).

Funding for SDSS-III has been provided by the Alfred P. Sloan Foundation, the Participating Institutions, the National Science Foundation, and the U.S. Department of Energy Office of Science. The SDSS-III web site is http://www.sdss3.org/.
SDSS-III is managed by the Astrophysical Research Consortium for the Participating Institutions of the SDSS-III Collaboration including the University of Arizona, the Brazilian Participation Group, Brookhaven National Laboratory, Carnegie Mellon University, University of Florida, the French Participation Group, the German Participation Group, Harvard University, the Instituto de Astrofisica de Canarias, the Michigan State/Notre Dame/JINA Participation Group, Johns Hopkins University, Lawrence Berkeley National Laboratory, Max Planck Institute for Astrophysics, Max Planck Institute for Extraterrestrial Physics, New Mexico State University, New York University, Ohio State University, Pennsylvania State University, University of Portsmouth, Princeton University, the Spanish Participation Group, University of Tokyo, University of Utah, Vanderbilt University, University of Virginia, University of Washington, and Yale University.

This publication makes use of data products from the Wide-field Infrared Survey Explorer, which is a joint project of the University of California, Los Angeles, and the Jet Propulsion Laboratory/California Institute of Technology, funded by the National Aeronautics and Space Administration. This research has made use of the NASA/ IPAC Infrared Science Archive, which is operated by the Jet Propulsion Laboratory, California Institute of Technology, under contract with the National Aeronautics and Space Administration.

This work was financially supported in part by the Japan Society for the Promotion of Science (JSPS; TN: grant No. 25707010, 16H01101, and 16H03958). T. Miyaji is supported by UNAM-DGAPA Grant PAPIIT IN104216 and CONACyT Grant Investigaci\'on B\'asica 252531. KM is financially supported by the JSPS through the JSPS Research Fellowship for research abroad. NK acknowledges the financial support of Grant-in-Aidfor Young Scientists (B:16K17670).
This research used the facilities of the Canadian Astronomy Data Centre operated by the National Research Council of Canada with the support of the Canadian Space Agency. This research made use of Astropy, a community-developed core Python package for Astronomy (\citealt{2013A&A...558A..33A}). Data analysis was in part carried out on common use data analysis computer system at the Astronomy Data Center, ADC, of the National Astronomical Observatory of Japan.
\bibliography{adssample}

\end{document}